\documentstyle[amsfonts,amssymb,11pt,epsfig,diagram]{article}

\makeatletter
\@addtoreset{equation}{section}

\newcommand{\R}{{\Bbb R}}
\newcommand{\C}{{\Bbb C}}
\newcommand{\Z}{{\Bbb Z}}

\def\SL{{\rm SL}}
\def\H{{{\bf H}^3}}

\begin{document}

\begin{titlepage}

\thispagestyle{empty}

\title{Analytic Continuation for\\
Asymptotically AdS 3D Gravity}

\author{
{\bf Kirill Krasnov}\thanks{{\tt
krasnov@cosmic.physics.ucsb.edu}}
\\
{\it Department of Physics}\\
{\it University of California, Santa Barbara, CA 93106}}

\date{\normalsize November, 2001}
\maketitle

\begin{abstract}
\normalsize We have previously proposed that asymptotically AdS 3D
wormholes and black holes can be analytically continued to the
Euclidean signature. The analytic continuation procedure was
described for non-rotating spacetimes, for which a plane $t=0$ of
time symmetry exists. The resulting Euclidean manifolds turned out
to be handlebodies whose boundary is the Schottky double of the
geometry of the $t=0$ plane. In the present paper we generalize
this analytic continuation map to the case of rotating wormholes.
The Euclidean manifolds we obtain are quotients of the hyperbolic
space by a certain quasi-Fuchsian group. The group is the
Fenchel-Nielsen deformation of the group of the non-rotating
spacetime. The angular velocity of an asymptotic region is shown
to be related to the Fenchel-Nielsen twist. This solves the
problem of classification of rotating black holes and wormholes in
2+1 dimensions: the spacetimes are parametrized by the moduli of
the boundary of the corresponding Euclidean spaces. We also
comment on the thermodynamics of the wormhole spacetimes.
\end{abstract}

\end{titlepage}

%\tableofcontents

\section{Introduction}
\label{sec:intr}

The black and wormholes \cite{Brill,Rot} of asymptotically AdS
3D gravity are of great interest, not only because this is
a large, non-trivial, yet explicitly constructible class
of spacetimes. These spacetimes can also be expected to
play a fundamental role in the would be theory of
quantum 3D gravity, as some of its building blocks. In this
paper we study an analytic continuation procedure that gives
an explicit construction of a Euclidean asymptotically AdS
manifold corresponding to a black or wormhole. Such
a procedure was initially described in \cite{Riemann}
for the case of non-rotating spacetimes. In the present
paper we generalize it to the rotating case.

There are several motivations for the analytic continuation
construction. The most obvious one is the black hole
thermodynamics. One obtains the thermodynamic partition function
of a black hole as the path integral over the Euclidean metrics.
As we shall see, the analytic continuation also helps to classify
the black and wormhole spacetimes. Finally, there are also reasons
to expect that an analytic continuation procedure will play an
important role in a construction of the Lorentzian quantum theory.

\begin{figure}
\centerline{\hbox{\epsfig{figure=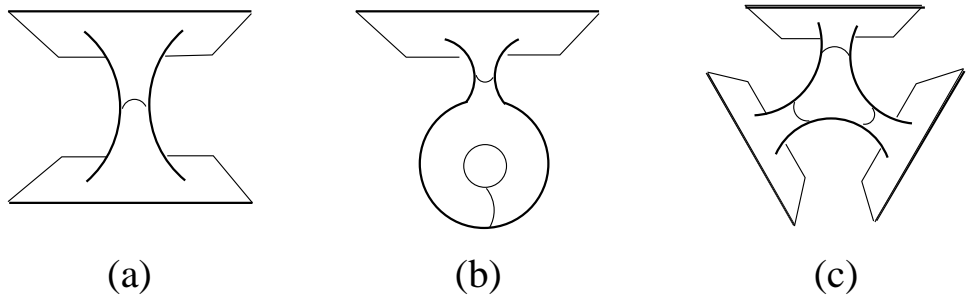,height=1.5in}}}
\caption{The geometry of the time symmetry plane for: (a) BTZ BH
(b) a single asymptotic region wormhole (c) three asymptotic
regions black hole.}
\label{fig:bh}
\end{figure}

Before we describe the main idea of our procedure, let us remind
the reader some basic facts about black holes and wormholes in 2+1
dimensions. These spacetimes generalize the BTZ BH \cite{BTZ}. It
is easiest to describe the non-rotating case. Then there is a
plane of time symmetry, which we shall call the $t=0$ surface. The
geometry of a spacetime is most conveniently described by looking
at this surface. Thus, a non-rotating BTZ BH has two asymptotic
regions and its $t=0$ plane has the geometry of an infinite throat
connecting the two regions, see Fig.~\ref{fig:bh}(a). More
interesting spacetimes can be constructed, see \cite{Brill}. For
example, as the $t=0$ surface geometry one can have a single
asymptotic region glued to a torus, see Fig.~\ref{fig:bh}(b). Or,
instead, one can have three asymptotic regions, see
Fig.~\ref{fig:bh}(c). The geometries described are just simplest
examples. More generally one can have any number of asymptotic
regions connected by a system of throats, with possibly handles
hidden behind the horizons. The corresponding spacetimes are
constructed from AdS${}_3$ by identifying points with respect to
the action of some discrete group of isometries, see more on this
below.

In \cite{Riemann} we have proposed that there is a natural way to
analytically continue the described spacetimes to the Euclidean
signature, at least in the non-rotating case. The idea in
\cite{Riemann} was as follows. The group $\Gamma$ used to obtain
a wormhole is a subgroup of a certain diagonal $\SL(2,\R)\subset
\SL(2,\R)\times\SL(2,\R)$. But such $\Gamma$ can also be
thought of as a subgroup of
$\SL(2,\C)$, which is the isometry group of the Euclidean AdS.
Thus, one can use {\it the same} identifications, but now
thought of as transformations of the Euclidean AdS, to obtain
a constant negative curvature 3D manifold. The corresponding
Euclidean manifolds turned out to be handlebodies. For examples,
both the single asymptotic region wormhole Fig.~\ref{fig:bh}(b)
and the three asymptotic regions black hole Fig.~\ref{fig:bh}(c)
correspond in the Euclidean signature to solid 2-handled spheres.

Before we explain how this procedure can be generalized to the
case of rotating spacetimes, let us make few comments.
First, we have to emphasize that the construction proposed
in this paper is not the standard analytic continuation procedure in
which one continues an appropriate time coordinate. Our construction
only works in the case of 2+1 dimensions. It is in this case
that there are no local degrees of freedom, and all spacetimes
are locally indistinguishable from the maximally symmetric one,
and thus obtainable from it by discrete identifications. This
raises a possibility of ``analytically continuing'' the discrete groups one
uses, not the time coordinate. Second, let us compare our
prescription to the more standard continuation of the time
coordinate. When one has a global time KVF in a spacetime, as is
the case for the BTZ BH, the result of our procedure coincides with what
one gets by continuing the time coordinate. In the case of
a spacetime with several asymptotic regions no global time
KVF is present, see \cite{Brill} for a discussion of this.
However, each asymptotic region is identical
to the asymptotic region of BTZ BH of some mass and angular
momentum. Thus, one could attempt to analytically continue
the metric in each asymptotic region separately. As is
usual with the analytic continuation of BH spacetimes, the
region behind the horizons ``disappears'', and the resulting
Euclidean space is a number of disconnected copies
of the Euclidean BTZ BH, that is solid tori. In contrast,
what one gets as a result of our procedure is a connected
space, whose boundary is some connected Riemann surface.
Which of these two analytic continuations is the ``correct'' one
depends on how one wants to use the resulting Euclidean spaces.
We make more comments on this in the sequel. For now we note
that from the point of view of holography it is natural to have the boundary
of the Euclidean space a connected surface. Indeed, one expects
to have non-zero correlations between the asymptotic
regions of a spacetime, so that a thermal state comes
from tracing a pure state over the degrees of freedom of
all asymptotic regions but one. In the framework of holography
such correlations are naturally incorporated via the
field theory living on a connected boundary of the
Euclidean space. For us this serves as a strong motivation
to prefer the procedure of continuing the discrete groups
to the procedure of analytically continuing each of the asymptotic
regions separately.

Let us now describe the main idea of our procedure. The idea is to
``analytically continue'' the discrete group
$\Gamma\in\SL(2,\R)\times\SL(2,\R)$ to a subgroup
$\Sigma\in\SL(2,\C)$. Then a spacetime that is obtained from the
Lorentzian AdS${}_3$ by identifying points with respect to the
action of $\Gamma$ corresponds to a 3D hyperbolic space
$\H/\Sigma$. To see why such an analytic continuation is possible,
and, in fact, natural, let us recall that the Lie algebra of
$\SL(2,\C)$ can be split into the selfdual and anti-selfdual
commuting parts. Then any $\SL(2,\C)$ transformation can be
represented as a composition of two transformations. In fact, this
is the usual construction of the vector representation of the
Lorentz group out of the fundamental and anti-fundamental one. One
constructs a $2\times 2$ matrix $\bf S$ that realizes the
fundamental representation. The matrix $\bf S$ is exactly the
``selfdual part'' of a Lorentz transformation. The vector
representation is then ${\bf x}\to {\bf S x S^\dagger}$, where
$\bf x$ is a unitary $2\times 2$ matrix constructed out of the
coordinates of $\R^{1,3}$, see more on this in the Appendix. Any
$\SL(2,\C)$ transformation can be represented this way. This means
that the isometry group in the Euclidean case can be given the
same product structure as in the Lorentzian signature. One can
thus analytically continue each of $\SL(2,\R)$'s of the Lorentzian
group of isometries into a selfdual (anti-selfdual) part of
$\SL(2,\C)$.

This procedure can actually be carried out for simple spacetimes,
like that of BTZ black hole, see the main text. However, for more
complicated black holes the procedure becomes ineffective in that
it is rather hard to see what the final Euclidean space is. In the
present paper we also develop a more effective description using
quasi-conformal mappings and quasi-Fuchsian groups. The main idea
is that the group $\Sigma$ can be obtained as a certain
deformation of the Fuchsian group $\Gamma$ of the non-rotating
spacetime. Namely, we construct $\Sigma$ as the so-called
Fenchel-Nielsen deformation \cite{Wol-1} of the non-rotating group
$\Gamma$. We show that, at least infinitesimally, the angular
velocity is related to the Fenchel-Nielsen twist. However, to
explain this result we will need to introduce a large amount of
background material, which we do in the main text.

In this paper we do not consider an important problem of
analytically continuing the Green's functions. This
problem has to be faced if one wants to do quantum field
theory in black hole spacetimes.

The paper is organized as follows. In the next section
we remind the reader how the non-rotating spacetimes
are constructed, and explain how they can be analytically
continued to the Euclidean signature. Section \ref{sec:rot-lor}
gives some necessary background on the rotating spacetimes.
Finally, the analytic continuation for the rotating case
is described in Section \ref{sec:rot-eucl}. we conclude with
a brief discussion.

\section{Non-rotating wormholes and their analytic continuation}
\label{sec:review}

We start by reviewing how non-rotating black and wormholes are
constructed \cite{Brill} and how they can be analytically continued
to the Euclidean signature \cite{Riemann}. Most of this section is
a review of the material contained in the above references. The
only new part is contained at the end: we explain why the
geometry of the boundary of the Euclidean space is that of
the Schottky double of the time symmetry plane geometry.

Let us start by reminding the reader some very basic facts about
the Lorentzian AdS${}_3$, out of which more complicated spaces
will be obtained by identifications of points. The spacetime is
best viewed as the interior of an infinite cylinder. The cylinder
itself is the conformal boundary $\cal I$ of the spacetime. It is
timelike, unlike the null conformal boundary of an asymptotically
flat spacetime. All light rays propagating inside AdS start and
end on $\cal I$. In this picture the constant time slices are
copies of the Poincare (unit) disc. The unit disc is isometric to
the upper half plane $\bf H$; we shall use both models. The
isometry group of the Lorentzian signature AdS${}_3$ is
$\SL(2,\R)\times\SL(2,\R)$. The spacetimes itself can be viewed as
the $\SL(2,\R)$ group manifold, and the isometry group action is
just the left and right multiplication. More details on AdS${}_3$
are given in the Appendix.

Let us now turn to the black hole spacetimes. The description of
the non-rotating black holes is greatly facilitated by the fact
that there is a surface $t=0$ of time symmetry. This surface is
preserved by the discrete group $\Gamma$ one uses to identify
points. Thus, $\Gamma$ is actually a subgroup of the group
$\SL(2,\R)\subset\SL(2,\R)\times\SL(2,\R)$ that fixes the $t=0$
plane. This ``diagonal'' $\SL(2,\R)$ consists of transformations
of the form ${\bf x}\to g {\bf x} g^T, g\in\SL(2,\R)$, where we
imply the model of AdS${}_3$ as the $\SL(2,\R)$ group manifold,
see the Appendix. Note that this is not the usual diagonal
$\SL(2,\R)$ consisting of transformations ${\bf x}\to g{\bf
x}g^{-1}, g\in\SL(2,\R)$, which fixes the origin of AdS${}_3$.
Transformations fixing the $t=0$ plane act on it by isometries.
Thus, the geometry of the surface $t=0$ is that of the quotient of
the unit disc by the action of $\Gamma\subset\SL(2,\R)$. Such 2D
geometries were an object of an active study by mathematicians for
the last hundred years. Once the geometry of the $t=0$ plane is
understood one just ``evolves'' the identifications in time to
obtain a spacetime, see \cite{Brill}.

\begin{figure}
\centerline{\hbox{\epsfig{figure=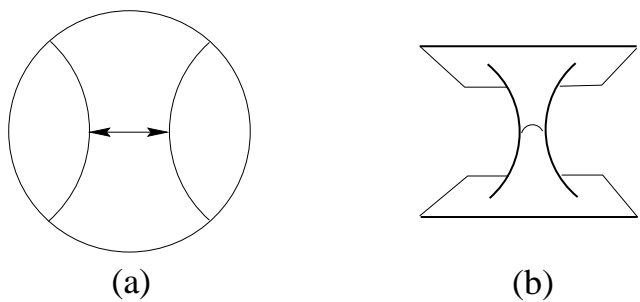,height=1.5in}}}
\caption{BTZ black hole: the geometry of the time symmetry surface.}
\label{fig:btz}
\end{figure}

Let us see how this works on examples. Consider first the case of the
non-rotating BTZ BH. In this case the discrete group is generated by a single
hyperbolic element. Its action on the $t=0$ plane can be
understood by finding the so-called fundamental region. The
fundamental region $D$ of the unit disc $\bf H$ for group $\Gamma$
is such that any point on $\bf H$ can be obtained as an image of a point
in $D$ under a transformation from $\Gamma$, and such that no
two points of $D$ (except on its boundary) are related. In the case
of $\Gamma$ generated by a single element $\gamma$ the fundamental region
is that between two geodesics on $\H$ mapped into one another
by the generator $\gamma$, see Fig.~\ref{fig:btz}(a). It is clear
that the quotient space has the topology of the
$S^1\times\R$ wormhole with two asymptotic regions,
each having the topology of $S^1$, see Fig. \ref{fig:btz}(b).
The BTZ angular coordinate runs from one
geodesics to the other. The distance between the two
geodesics measured along their common normal
is precisely the horizon circumference.

\begin{figure}

\unitlength 0.800mm
\linethickness{0.4pt}
\begin{picture}(122.87,72.61)(0,0)%74.61)
\thicklines

\bezier{40}(116.60,14.31)(116.60,11.06)(108.76,9.43)
\bezier{40}(108.76,9.43)(100.93,7.81)(93.09,9.43)
\bezier{40}(93.09,9.43)(85.26,11.06)(85.26,14.31)
\bezier{40}(116.60,42.15)(116.60,39.47)(108.76,38.14)
\bezier{40}(108.76,38.14)(100.93,36.81)(93.09,38.14)
\bezier{40}(93.09,38.14)(85.26,39.47)(85.26,42.15)
\bezier{16}(105.63,37.75)(104.67,38.56)(108,40)
\bezier{22}(108,40)(111.42,40.5)(113.46,39.37)
\bezier{16}(96.23,46.54)(97.18,45.73)(94.66,44.93)
\bezier{22}(94.66,44.93)(91.44,44.23)(88.39,44.93)
\bezier{40}(116.60,69.99)(116.60,72.30)(108.76,73.45)
\bezier{40}(108.76,73.45)(100.93,74.61)(93.09,73.45)
\bezier{40}(93.09,73.45)(85.26,72.30)(85.26,69.99)
\bezier{40}(116.60,69.99)(116.60,67.68)(108.76,66.53)
\bezier{40}(108.76,66.53)(100.93,65.37)(93.09,66.53)
\bezier{40}(93.09,66.53)(85.26,67.68)(85.26,69.99)
\put(85.26,14.31){\line(0,1){55.68}}
\put(116.60,14.31){\line(0,1){55.68}}
\thicklines
\multiput(88.78,67.48)(0.50,0.12){46}{\line(1,0){0.50}}

\multiput(88.39,10.96)(0.42,0.12){58}{\line(1,0){0.42}}

\bezier{260}(88.78,67.48)(122.87,36.21)(88.39,10.96)
\bezier{74}(111.90,72.77)(117.38,64.60)(116.60,53.28)
\bezier{64}(115.03,31.57)(111.90,39.55)(115.03,47.72)
\bezier{40}(115.03,47.72)(116.60,53.28)(116.60,53.28)
\bezier{44}(115.03,31.57)(116.60,27.11)(116.60,25.44)
\bezier{22}(116.60,25.44)(116.60,20.99)(112.68,17.93)
%RH edge
\bezier{17}(111.90,72.77)(103.02,65.16)(98.58,56.63)
\bezier{114}(98.58,56.63)(91.53,42.89)(103.28,28.23)

\bezier{12}(103.28,28.23)(106.41,23.40)(112.68,17.93)
\bezier{74}(88.39,10.96)(84.47,19.69)(85.26,31.01)
\bezier{64}(86.82,53.84)(89.96,44.75)(86.82,36.58)
\bezier{40}(86.82,36.58)(85.26,31.01)(85.26,31.01)
\bezier{44}(86.82,53.84)(85.26,57.18)(85.26,59.97)
\bezier{22}(85.26,60.52)(85.26,63.31)(88.78,67.48)
\thinlines

\put(117.38,42.15){\makebox(0,0)[lc]{$t=0$}}
\bezier{40}(116.60,14.31)(116.60,16.25)(112.68,17.93)
\bezier{10}(112.68,17.93)(104.85,20.43)(98.58,19.87)
\bezier{8}(98.58,19.87)(90.35,19.32)(86.43,16.53)
\bezier{10}(86.43,16.53)(85.26,15.59)(85.26,14.31)
%lowest circle
\bezier{16}(85.26,42.15)(84.94,43.71)(88.39,44.97)
\bezier{5}(88.39,44.97)(91.94,46.23)(95.91,46.60)
\bezier{40}(95.91,46.60)(99.89,46.98)(103.96,46.75)
\bezier{7}(103.96,46.75)(110.02,46.31)(114.09,44.67)
\bezier{12}(114.09,44.67)(116.18,44.00)(116.60,42.15)
%middle circle

%\put(101.00,5.00){\makebox(0,0)[cc]{(b)}}
\bezier{92}(89.00,67.00)(88.61,57.78)(94.00,45.00)
\bezier{140}(89.00,67.00)(103.89,56.39)(108.00,40.00)
%horizon
\put(94.00,44.80){\line(3,-1){11.50}}
\put(93.00,48.00){\line(3,-1){12.00}}
\put(92.00,51.00){\line(3,-1){12.00}}
\put(91.00,54.00){\line(3,-1){11.80}}
\put(90.00,57.00){\line(3,-1){11.00}}
\put(89.50,60.00){\line(3,-1){9.00}}
\put(89.00,62.50){\line(3,-1){7.00}}

\multiput(89,65)(0.42,-0.13){8}{\line(1,0){0.42}}
\bezier{5}(105.4,41.00)(106.5,40.50)(107.99,40.00)
\bezier{3}(105.3,44.00)(106.1,43.65)(106.94,43.30)
\bezier{3}(103.94,47.00)(104.9,46.75)(105.78,46.50)
\bezier{3}(102.8,50.00)(103.65,49.75)(104.5,49.50)
\bezier{2}(101.28,53.17)(101.7,53.08)(102.11,53.00)
\bezier{2}(98.44,56.83)(99.22,56.66)(99.94,56.50)
\bezier{2}(96.44,60.00)(97.72,59.91)(97.00,59.82)
\bezier{144}(108.00,40.00)(117.78,61.11)(112.00,73.00)
\end{picture}

\caption{BTZ black hole: the spacetime picture. Two
asymptotic regions are shown: they are parts of the boundary cylinder
lying between the timelike geodesics. The future event horizon,
which is the boundary of the past of the asymptotic infinity,
is shown. Note that it intersects the initial slice along the
minimal line connecting the two geodesics bounding the fundamental
region.}
\label{fig:btz-st}
\end{figure}
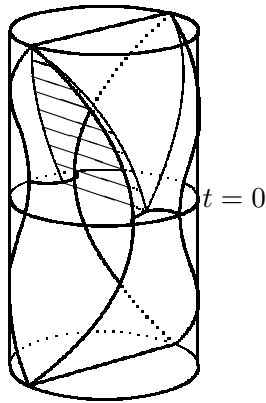

One can obtain the spacetime geometry of this black hole
by ``evolving'' in time the $t=0$ slice geometry. In this evolution,
the two geodesics on the initial slice
evolve into geodesic surfaces in the spacetime, which are then
to be identified. The region between them,
after identifications, is the BTZ black hole spacetime,
see Fig. \ref{fig:btz-st}. Because timelike geodesics in
AdS are ``attracted'' to each other, the two geodesic
surfaces will finally cross. This is where the spacetime
``ends''. Note, however, that the time it takes for them
to meet each other is finite only in the global AdS
time coordinate. In BTZ time coordinate the corresponding
``singularities'' are at past and future infinity. The
asymptotic infinity of the BTZ black hole consists of
two regions on the AdS boundary cylinder. These are the
regions between the timelike geodesics on the boundary;
the geodesics are identified. One can now construct the
past of the asymptotic region to convince oneself that there
is a region in this spacetime that is causally disconnected
from the infinity, see Fig. \ref{fig:btz-st}.
This region is the black (white) hole, and its
boundary is the event horizon.

\begin{figure}
\centerline{\hbox{\epsfig{figure=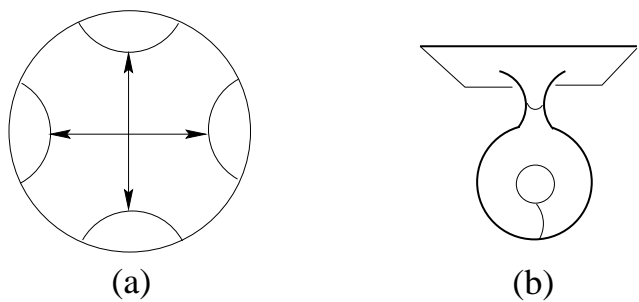,height=1.5in}}}
\bigskip
\caption{Initial slice geometry of the single asymptotic region
black hole with a torus wormhole inside the horizon}
\label{fig:wormhole}
\end{figure}

Let us now consider more complicated initial slice geometries.
We now consider the group $\Gamma$ to be generated by two
hyperbolic elements. For example, let the fundamental region be
the part of the unit disc between four geodesics, as in
Fig. \ref{fig:wormhole}(a). Let us identify these geodesics
cross-wise. It is straightforward to show that the resulting
geometry has only one asymptotic region, consisting of all
four parts of the infinity of the fundamental region. With
little more effort one can convince oneself that the resulting
geometry is one asymptotic region ``glued'' to a torus,
see Fig. \ref{fig:wormhole}(b). As far as the spacetime
obtained by evolving this geometry is concerned, one finds that this
is a single asymptotic region black hole, but the topology
{\it inside} the event horizon is now that of a torus.
See \cite{Brill} for more details on this spacetime.

\begin{figure}
\centerline{\hbox{\epsfig{figure=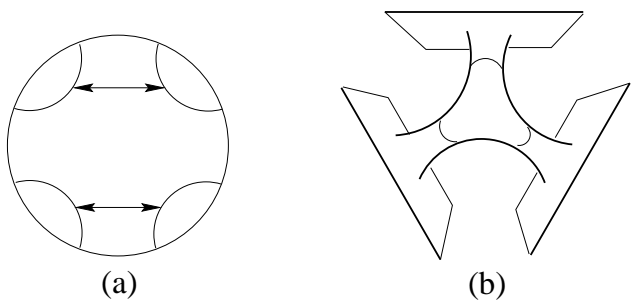,height=1.5in}}}
\caption{Initial slice geometry of the three asymptotic region black hole}
\label{fig:3bh}
\end{figure}

A group generated by two elements can also be used to obtain
a three asymptotic region black hole \cite{Brill}. The fundamental
region on the $t=0$ plane is again the region bounded by four
geodesics. They are, however, now identified side-wise,
see Fig. \ref{fig:3bh}(a). One can clearly see that the
initial slice geometry has three asymptotic regions, as
in Fig. \ref{fig:3bh}(b). Evolving this, one gets a spacetime
with three asymptotic regions and corresponding event horizons.
See \cite{Brill} for more details.

Taking the group $\Gamma$ to be more complicated
one constructs a large class of spacetimes.
In particular, one can have a single asymptotic region
black hole with an arbitrary Riemann surface inside
the horizon. More generally, one can have a black hole
with any number of asymptotic regions, and with any number of
handles hidden behind the horizon(s).

Let us now turn to the procedure of analytic continuation. The
basic idea is, instead of analytically continuing the metric in
some time coordinate, produce a space by identifying points in the
Euclidean AdS${}_3$ {\it using the same group} $\Gamma$. We recall
that the group of isometries of the Euclidean AdS${}_3$
(=hyperbolic space $\H$) is $\SL(2,\C)$, see the Appendix for more
detail. However, $\SL(2,\R)$ is naturally a subgroup of
$\SL(2,\C)$, thus $\Gamma$ acts on $\H$ and this action can be
used to obtain a quotient space. To see what this quotient space
is let us give another, equivalent description of the continuation
map. Let us take a section of the Poincare ball (a model for $\H$)
by a plane passing through the center of the ball. The
intersection of the ball with the plane is a unit disc $\bf H$.
Let us call this $t=0$ plane and do on it the same identifications
as we do on the time symmetry plane of the spacetime to be
analytically continued. Let us then ``evolve'' these
identifications, but now in the Euclidean time. To do this one
just constructs geodesic surfaces intersecting the $t=0$ plane
orthogonally along the geodesics bounding the fundamental region.
The geodesic surfaces in $\H$ are hemispheres; they are to be
identified.

\begin{figure}
\centerline{\hbox{\epsfig{figure=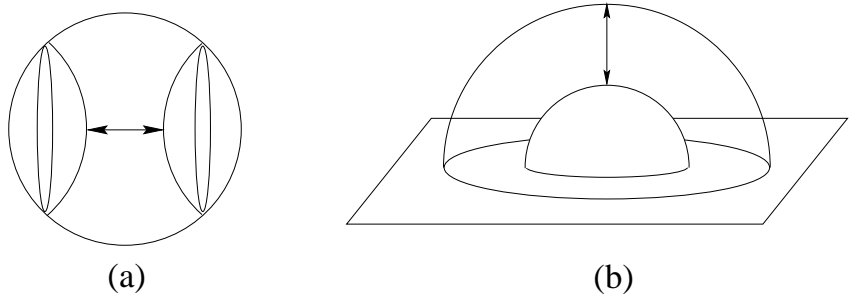,height=1.5in}}}
\caption{Euclidean BTZ black hole}
\label{fig:btz-eucl}
\end{figure}

Let us see how this works for the simplest case of the BTZ black hole.
Thus, we require that the geometry of the $t=0$ slice of the
unit ball is the same as the geometry of the
$t=0$ slice of BTZ black hole, see Fig. \ref{fig:btz}. We then
have to build geodesic surfaces
above and below the two geodesics on the $t=0$ plane, see
Fig. \ref{fig:btz-eucl}(a). The Euclidean
BTZ black hole is then the region between these hemispheres;
the hemispheres themselves are identified. It is clear that
the space obtained is a solid torus, its conformal boundary
being a torus. It is often more convenient to
work with another model for the same space,
that using the upper half space. The interior of the Poincare
ball can be isometrically mapped into the upper half-space.
The boundary sphere goes under this map into the $x-y$ plane.
In the case of $\Gamma$ generated by a single generator one
can always put its fixed points to $0,\infty$, so that
the picture of the Euclidean BTZ BH becomes that in
Fig.~\ref{fig:btz-eucl}(b). It is important that
using our procedure we have arrived at the same space
as is the one obtained by the usual analytic continuation
in the time coordinate, see \cite{Carlip-T}.

\begin{figure}
\centerline{\hbox{\epsfig{figure=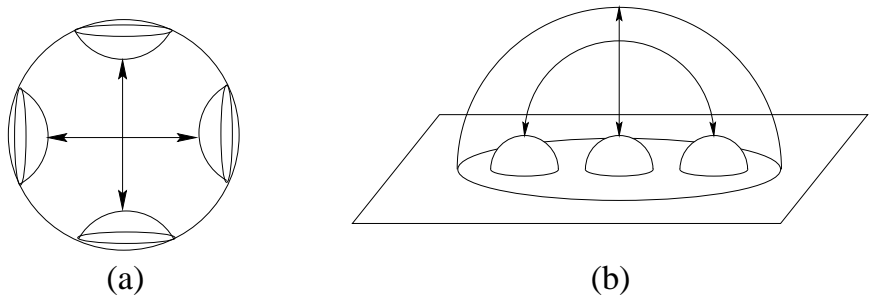,height=1.5in}}}
\caption{Euclidean single asymptotic region black hole with a torus inside}
\label{fig:wormhole-eucl}
\end{figure}

Let us consider another example. We now want to construct the
Euclidean version of the single asymptotic region black hole
with a torus inside the horizon. The procedure is the same:
we require a slice of the unit ball to have the same geometry
as the $t=0$ slice of the black hole. This gives us four
hemispheres inside the unit ball; the Euclidean space is the
region between them and they are to be identified cross-wise,
see Fig. \ref{fig:wormhole-eucl}(a). One sees that the
Euclidean space is a solid 2-handled sphere. One can again
map the whole configuration into the upper half-space,
see Fig. \ref{fig:wormhole-eucl}(b).

One can do a similar analysis for the three asymptotic region
black hole (one also gets a solid two-handled sphere),
and for any other of the non-rotating black holes
of \cite{Brill}. In all cases the pattern is the same: one
requires the $t=0$ slice geometry to be the same also in the
Euclidean case, and this determines the Euclidean geometry
completely. The Euclidean spaces one gets are handlebodies.

We would now like to understand in more detail a relation between
the geometry of the time symmetry plane and the conformal boundary
of the Euclidean space. As we shall explain, the later is
essentially the so-called Schottky double of the former. First,
we have to explain what the Schottky double is. This concept
plays an important role in boundary conformal field theory,
see, e.g., a recent review \cite{Schw}. Given a Riemann surface $X$,
closed or with a boundary, its Schottky double is another Riemann
surface $\tilde{X}$, not necessarily connected, out
of which the original surface can be obtained by
identifications: $X=\tilde{X}/\sigma$. Here $\sigma$
is an anti-holomorphic map of $\tilde{X}$ into itself.
For a surface $X$ without a boundary the Schottky double
$\tilde{X}$ is essentially given by two disconnected copies
of $X$, with all moduli replaced by their complex conjugates
in the second copy. For a surface with a boundary one takes
two copies of $X$ and glues them along the boundary to obtain
a connected surface. The anti-holomorphic map $\sigma$ fixes
the pre-image of the boundary of $X$ on $\tilde{X}$.

\begin{figure}
\centerline{\hbox{\epsfig{figure=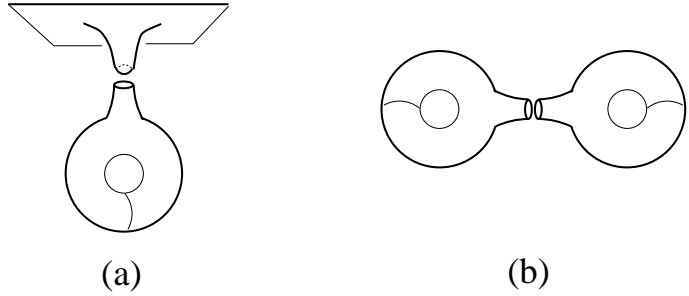,height=1.5in}}}
\caption{(a) Disconnecting the asymptotic regions one gets a
Riemann surface $X$ with boundaries; (b) One glues two copies of
$X$ to obtain the Schottky double $\tilde{X}$.}
\label{fig:disconnect}
\end{figure}

Let us now use this concept in our story. The $t=0$ plane
geometries one gets can be thought of in two ways. First, one can
view them as throats connecting asymptotic regions. However, one
can also disconnect the asymptotic regions along the minimal
geodesics (horizons). What one gets is a Riemann surface $X$ with
one circular boundary for every asymptotic region, see
Fig.~\ref{fig:disconnect}(a). One takes two copies of this Riemann
surface, and glues them along the boundary circles to get a close
surface --the Schottky double $\tilde{X}$, see
Fig.~\ref{fig:disconnect}(b).

It is not hard to convince oneself that the geometry of the
boundary of a Euclidean space obtained via our analytic
continuation prescription is that of the Schottky double of the
$t=0$ geometry. This is related to the fact that the Riemann
surface one obtains is uniformized by the complex plane, which is
the so-called uniformization by Schottky groups. Let us first
recall some basic information about the Schottky groups, see,
e.g., \cite{Hyperb} as a reference. A Schottky group $\Sigma$ is a
discrete subgroup of $\SL(2,\C)$, freely (that is, no relations)
generated by a number $g$ of loxodromic (that is ${\rm Tr}(L_i)
\notin [0,2]$) generators $L_1,\ldots,L_g\in\SL(2,\C)$. The
Schottky group $\Sigma$ acts by conformal transformations on the
complex plane $\C$. Let us denote by $\cal C$ the complement of
the set of fixed points of this action. As is not hard to convince
oneself, the quotient ${\cal C}/\Sigma$ is a compact genus $g$
Riemann surface. A Riemann surface obtained from the complex plane
by identifications from a Schottky group is called uniformized via
Schottky. This is a uniformization different from the usual
Fuchsian one that uses the hyperbolic plane. We have already
encountered surfaces uniformized by Schottky groups. The
boundaries of our Euclidean spaces were obtained exactly this way.
It is only that we considered Schottky groups that are real, that
is, subgroups of $\SL(2,\R)$. This is related to the fact that we
have so far only considered non-rotating spacetimes. As we explain
later, inclusion of rotation would amount to considering general
Schottky groups.

\begin{figure}
\centerline{\hbox{\epsfig{figure=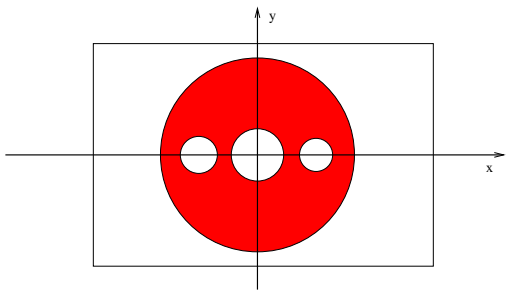,height=1.5in}}}
\caption{The fundamental domain for the Schottky uniformization
of a genus 2 surface.}
\label{fig:Schottky}
\end{figure}

Let us illustrate all this on an example, and also explain why the
Euclidean boundary is the Schottky double. Consider the single
asymptotic region wormhole. The $t=0$ plane geometry is obtained
as a quotient with respect to a group $\Gamma\subset\SL(2,\R)$
generated by two elements. The same group, thought of as a
subgroup of $\SL(2,\C)$ acts in $\H$. In particular, it acts on
the boundary of $\H$, that is the complex plane, by fractional
linear transformations. The boundary of the Euclidean space is
then obtained as the quotient of the complex plane by this action.
The fundamental region for this action is shown in
Fig.~\ref{fig:Schottky}. Since all generators are in $\SL(2,\R)$,
their fixed points are located on the real axes, and so are the
centers of the circles bounding the fundamental region. Removing
the circles one gets a sphere with four holes. Identifying their
boundaries one gets a genus 2 surface --our Euclidean boundary.
One can also obtain the $t=0$ plane geometry from the same
picture. One should just take the upper half-plane, with four
semi-circles bounding the fundamental region for the action of
$\Gamma$ in $\bf H$. We see that the Euclidean boundary can be
thought of as made of two copies of the $t=0$ plane. This is why
the boundary of the Euclidean space is the Schottky double of the
time symmetry surface geometry. Let us note that the two copies
needed to obtain the Euclidean boundary are exactly the same.
Indeed, the configuration of circles in Fig.~\ref{fig:Schottky} is
invariant under the reflection on the real line. This reflection
is exactly the anti-holomorphic map $\sigma$ that is part of the
definition of the Schottky double. Below we shall see what
$\sigma$ is for the rotating spacetimes.

Having explained why the boundary of the Euclidean space is
the Schottky double of the initial slice geometry, let use this
fact to obtain a simple relation between the number of asymptotic
regions $K$, the number of handles $G$ behind the horizon, and the genus
of the Euclidean boundary $g$. As is not hard to see:
\begin{equation}
g = 2G + K - 1.
\end{equation}
This implies that, given a genus $g>1$ of the Euclidean boundary,
the ``internal'' genus $G$ can take values $0,\ldots,g/2$, for
$g$ even, and values $0,\ldots,(g-1)/2$ for $g$ odd. Thus,
for large genus, there are of order $g/2$ possible types of non-rotating
spacetimes that lead to the genus $g$ for their Euclidean boundary.

%\eject

\section{Rotating wormholes}
\label{sec:rot-lor}

To generalize the proposed analytic continuation procedure to
rotating spacetimes we first need to understand how to describe
the later. In this section we review some necessary facts. Our
reference here is \cite{Rot}, see also \cite{Brill-new}. Most
of our material here is from \cite{Rot}. However, we present a
new formula for the angular momentum at the end of the section.
We also give a correct expression for the angular momentum
of a single asymptotic region wormhole considered in \cite{Rot}.
The result for $\Omega$ in this reference is incorrect.

The main difference with the non-rotating case is that there is anymore
no plane of time symmetry. Thus, one needs a qualitatively new
way of describing the spacetime. The idea of \cite{Rot} was
to unravel the spacetime structure by considering the action of
the discrete group at the boundary, instead of thinking about the
action at the time symmetry plane that is no longer available.

\begin{figure}
\centerline{\hbox{\epsfig{figure=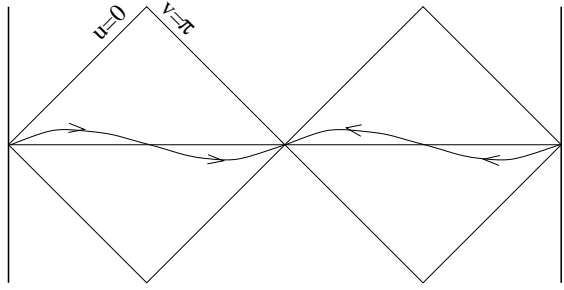,height=1.5in}}}
\caption{The picture of the conformal boundary $\cal I$ for a
rotating BTZ BH. There are two fixed points, and the region
where the VF generating rotations is spacelike is bounded by
the null lines passing through the fixed points. One of the
integral lines of $\xi_{rot}$ is shown.}
\label{fig:btz-rot}
\end{figure}

Let us now consider the rotating
BTZ BH. It is obtained from AdS${}_3$ identifying with
respect to transformations generated by $\gamma=e^{\xi_{rot}}$.
We choose $\xi_{rot}$ as in \cite{Rot}:
\begin{equation}\label{btz-rot}
\xi_{rot}=a J_{XU} +b J_{YV} =
-J_1(a+b)-\tilde{J}_1(a-b) =
-(a+b)\sin{u} \partial_u -(a-b)\sin{v} \partial_v.
\end{equation}
See the Appendix for a definition of the vector fields. As is
clear from this expression, the vector field $\xi_{rot}$ becomes
null along the lines $u=0,\pi, v=0,\pi$. The fixed points of
$\xi_{rot}$ are points where these null lines intersect: $u=v=0$
and $u=v=\pi$. One of the integral curves of $\xi_{rot}$ is shown
in Fig.~\ref{fig:btz-rot}. When $b=0$ this integral curve
coincides with the $t=0$ axes; this case corresponds to no
rotation.

To find the angular velocity of the corresponding spacetime
one proceeds in the following way, see \cite{Rot}. First, one finds
a new metric $d\hat{s}^2$ on $\cal I$, related to
$ds^2=-du dv$ by a conformal transformation. The requirement is
that $\xi_{rot}$ is a KVF of $d\hat{s}^2$ and that $||\xi_{rot}||^2=1$.
As is not hard to find, for $\xi_{rot}$ given by (\ref{btz-rot})
this metric is given by:
\begin{equation}\label{b-metric-btz}
d\hat{s}^2 = -{du dv\over (a^2-b^2)\sin{u}\sin{v}} = \omega^{-1} ds^2.
\end{equation}
Here we have introduced a notation $\omega$ for the conformal
factor relating $d\hat{s}^2$ and $ds^2$.
Next, one finds an asymptotic vector field $\xi_{time}$
generating the time translations. It is determined by the
conditions:
\begin{equation}
\xi_{time}\cdot \xi_{rot}=0, \qquad ||\xi_{time}||^2=-1.
\end{equation}
In our case
\begin{equation}
\xi_{time} = -(a+b)\sin{u} \partial_u +(a-b)\sin{v} \partial_v.
\end{equation}
The last step is to find the vector field that generates the
horizon. As is explained in \cite{Rot}, it is the VF that
is orthogonal to the surface $\omega=0$. We see that
\begin{equation}
\xi_{hor}=-\sin{u}\partial_u + \sin{v}\partial_v=
{a\over a^2-b^2}\left( \xi_{time} - \Omega \xi_{rot}\right),
\end{equation}
where
\begin{equation}\label{J}
\Omega={b\over a}.
\end{equation}
This is by definition a measure of how much the horizon rotates:
the angular velocity.

\begin{figure}
\centerline{\hbox{\epsfig{figure=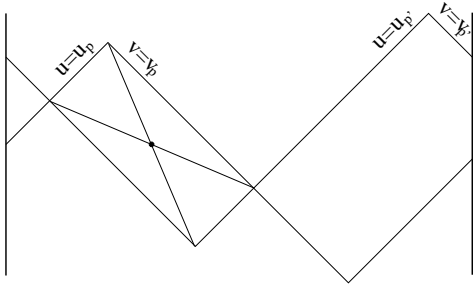,height=1.5in}}}
\caption{The conformal boundary $\cal I$ for a more general rotation
vector field. The ``center'' of any one of the two rectangles in
which the generator of rotations is spacelike plays a special role,
see the next section.}
\label{fig:general-rot}
\end{figure}

One can repeat this calculation for a more general VF. The
strategy is always the same. First, one finds fixed points of the
generator of rotations $\xi_{rot}$. The region on $\cal I$ where
$\xi_{rot}$ is spacelike is bounded by null geodesics passing
through the fixed points, see Fig.~\ref{fig:general-rot}. This
region is the covering of an asymptotic region of our spacetime.
One then finds a conformally rescaled metric $d\hat{s}^2$ and the
conformal factor $\omega$, finds a generator of time translations
$\xi_{time}$ and a horizon generator $\xi_{hor}$ orthogonal to the
$\omega=0$ surface. Finally, one should write $\xi_{hor}$ as a
linear combination of $\xi_{time}$ and $\xi_{rot}$, and read off
the angular velocity $\Omega$. In \cite{Rot} this procedure is
applied to an asymptotic region generated by the VF
\begin{equation}\label{general-rot}
\xi_{rot} = \sin{\left({u-u_p\over2}\right)}
\sin{\left({u-u_{p'}\over2}\right)} \partial_u -
k \sin{\left({v-v_p\over2}\right)}
\sin{\left({v-v_{p'}\over2}\right)} \partial_v.
\end{equation}
One finds:
\begin{equation}
\omega=-k \sin{\left({u-u_p\over2}\right)}
\sin{\left({u-u_{p'}\over2}\right)} \sin{\left({v-v_p\over2}\right)}
\sin{\left({v-v_{p'}\over2}\right)},
\end{equation}
and
\begin{equation}
\xi_{time} = \sin{\left({u-u_p\over2}\right)}
\sin{\left({u-u_{p'}\over2}\right)} \partial_u +
k \sin{\left({v-v_p\over2}\right)}
\sin{\left({v-v_{p'}\over2}\right)} \partial_v.
\end{equation}
This gives:
\begin{equation}\label{general-J}
\Omega= {\sin{\left({u_p-u_{p'}\over2}\right)}-k
\sin{\left({v_p-v_{p'}\over2}\right)}\over
\sin{\left({u_p-u_{p'}\over2}\right)}+k
\sin{\left({v_p-v_{p'}\over2}\right)}}.
\end{equation}
As a check of this formula let us note that it
becomes (\ref{J}) when $u_p=v_{p'}=0, u_{p'}=-\pi, v_p=\pi$. Then
for $k=(a-b)/(a+b)$ the rotation VF (\ref{general-rot}) reduces
(up to normalization) to (\ref{btz-rot}) and (\ref{general-J})
reduces to (\ref{J}).

As we will now show, the angular velocity (\ref{general-J}) can be
expressed in terms of the trace of the matrix
$\gamma=e^{\xi_{rot}}$. To this end, let us note that any
$\SL(2,\R)$ element can be parametrized as $\gamma=e^{s{\bf n}}$,
with ${\bf n} = \gamma_0 +\kappa\gamma(\theta)$, where
$\gamma(\theta)=\sin{\theta}\gamma_1+\cos{\theta}\gamma_2$.
Formula (\ref{geod}) of the Appendix shows that ${\rm Tr}\,\gamma
= 2{\rm cs}(s)$, where the later is given by (\ref{cs}). Let us
now take $\xi^L_{rot}=s{\bf n}$ as the ``left-moving'' part of our
rotation VF. Using the relation (\ref{J-gamma}) between gamma
matrices and the $J$-generators, and interpreting the generators
as those of the left-moving sector, we get, asymptotically:
\begin{equation}
s{\bf n} = - 2s\left(1 + \kappa(\sin{u}\sin{\theta}-\cos{u}\cos{\theta})
\right)\partial_u =
- 2s\left(1 - \kappa\cos{(u+\theta)}\right)\partial_u.
\end{equation}
Let us introduce: $1/\kappa=\cos{\delta}$. Then:
\begin{equation}
s{\bf n} = - 4s\kappa\sin{\left({u+\theta+\delta\over 2}\right)}
\sin{\left({u+\theta-\delta\over 2}\right)} \partial_u.
\end{equation}
This has the same form as the left-moving part of (\ref{general-rot}),
with $u_p=\delta-\theta, u_{p'}=-\delta-\theta$.
The expression (\ref{general-J}) involves
\begin{equation}
\sin{\left({u_p-u_{p'}\over 2}\right)} = \sin{\delta}=\sqrt{1-1/\kappa^2}.
\end{equation}
It is now easy to notice that
\begin{equation}
s\kappa\sin{\left({u_p-u_{p'}\over 2}\right)}=
{\rm Arccosh}\left({1\over 2}{\rm Tr}\,\gamma \right).
\end{equation}
Thus, the angular momentum (\ref{general-J}) can be expressed
through the traces of the left and right-moving
$\SL(2,\R)$ components of the isometry $\gamma=e^{\xi_{rot}}$. We have:
\begin{equation}\label{J*}
\Omega = {{\rm Arccosh}\left({1\over 2}{\rm Tr}\,\gamma^L \right)-
{\rm Arccosh}\left({1\over 2}{\rm Tr}\,\gamma^R \right)\over
{\rm Arccosh}\left({1\over 2}{\rm Tr}\,\gamma^L \right)+
{\rm Arccosh}\left({1\over 2}{\rm Tr}\,\gamma^R \right)}.
\end{equation}
This expression is often more convenient for calculating $\Omega$ than
(\ref{general-J}). The above formula for $\Omega$ is new.

\begin{figure}
\centerline{\hbox{\epsfig{figure=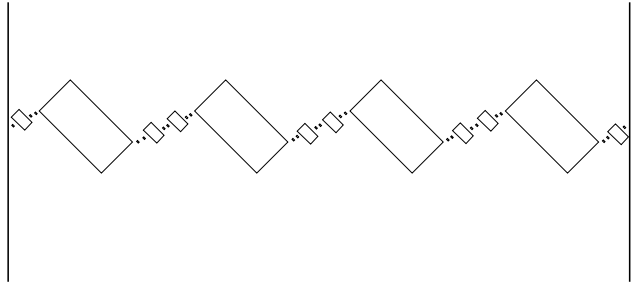,height=1.5in}}}
\caption{The conformal boundary $\cal I$ for a single asymptotic
region wormhole.}
\label{fig:wormhole-rot}
\end{figure}

As an example, let us apply (\ref{J*}) to a single asymptotic
region rotating wormhole \cite{Rot}. In this case the discrete
group is generated by two elements $\gamma_1, \gamma_2$. For a
certain symmetric configuration, see \cite{Rot}, the corresponding
vector fields have the following form:
\begin{eqnarray}\nonumber
\xi_1=-J_1(a+b)-\tilde{J}_1(a-b) =
-(a+b)\sin{u} \partial_u -(a-b)\sin{v} \partial_v, \\ \label{vf-wormhole}
\xi_2=J_2(a+b)-\tilde{J}_2(a-b) =
-(a+b)\cos{u} \partial_u +(a-b)\cos{v} \partial_v.
\end{eqnarray}
The boundary $\cal I$ has the structure depicted in
Fig.~\ref{fig:wormhole-rot}. One has an infinite number of
diamonds each being a copy of the covering space of the asymptotic
region. The generator of rotations in one of the copies of the
covering space is obtained as:
\begin{eqnarray*}\label{gamma}
\gamma=e^{\xi_{rot}}=[\gamma_1,\gamma_2]=\gamma_1\gamma_2
\gamma_1^{-1}\gamma_2^{-1}.
\end{eqnarray*}
To use (\ref{J*}) we need to find the left and right-moving parts
of $\gamma_1,\gamma_2$. Let us first find the left part. We have:
\begin{equation}
\gamma_1^L = e^{{1\over2}(a+b)\gamma_1}, \qquad
\gamma_2^L = e^{-{1\over 2}(a+b)\gamma_2}.
\end{equation}
Calculating the commutator we get:
\begin{equation}
{1\over2}{\rm Tr}\,\gamma^L = {1\over2}{\rm Tr}\,[\gamma^L_1,\gamma^L_2] =
1- 2\sinh^4{(a+b)/2}.
\end{equation}
The expression for the right part is obtained by
replacing $(a+b)\to(a-b)$. Note that the expression one gets
for $\Omega$ is different from the one given in \cite{Rot}.
Note that (\ref{gamma}) is the generator of rotations only in
one of the infinite number of copies of the covering space
of the asymptotic region. In other copies of the covering space
the generator is obtained from $\gamma$ by taking $g\gamma g^{-1}$
for all $g\in\Gamma$. Our expression (\ref{J*}) for the
angular velocity makes it clear that $\Omega$ is independent of
which copy of the covering space is used to determine it.
Indeed, $\Omega$ is invariant under $\gamma\to g\gamma g^{-1}$ since it
only depends on the trace of the left and right parts of $\gamma$,
which are both invariant under the conjugation.

\section{Analytic continuation: incorporation of rotation}
\label{sec:rot-eucl}

In this section we finally turn to the analytic continuation
procedure for rotating spacetimes. We start by showing how
one can get Euclidean manifolds by analytically continuing
the discrete groups.

\subsection{Continuation of the discrete groups}
\label{sec:groups}

As we have explained in the Introduction, one can analytically
continue a discrete subgroup $\Gamma\in\SL(2,\R)\times\SL(2,\R)$
into a discrete subgroup $\Sigma\in\SL(2,\C)$. Actually, the map
is only defined if one performes a continuation {\it back} from
Euclidean to the Lorentzian signature. Let us show how this works.
As we explained in the Appendix, any element of the Lorentz group
$\SL(2,\C)$ can be represented as a composition of commuting
``selfdual'' and ``anti-selfdual'' transformations. In the
spinorial realization of the Euclidean AdS${}_3$ by unimodular
unitary matrices this is just the vector representation ${\bf
x}\to{\bf S x S^\dagger}$, where $S$ is the matrix
\begin{equation}
{\bf S} = e^{{i\over 2}\sigma_i(\omega^i-i\nu^i)}
\end{equation}
realizing the fundamental representation. Let us analytically
continue $\bf S$ into an element of $\SL(2,\R)$ according to the
following rule:
\begin{eqnarray}\label{map}\nonumber
i\nu^i \to \nu^i, \\
i\sigma_1 \to \gamma_1, \qquad i\sigma_2 \to - \gamma_0,
\qquad i\sigma_3 \to \gamma_2.
\end{eqnarray}
One should apply this map to both $\bf S$ and ${\bf S}^\dagger$
and declare the obtained $\SL(2,\R)$ elements to be the left and
right parts of a Lorentzian isometry.

Let us now apply this continuation procedure to the Euclidean BTZ BH. In this
case, the generator is a boost along the vertical line plus some amount of
rotation:
\begin{equation}
L = e^{{i\over2} \sigma_3 (\phi+i\ln{|\lambda|})}.
\end{equation}
In order to analytically continue this generator to the Lorentzian
sector, we have to represent it as a sum of two commuting vector
fields and then analytically continue each of them. As the result,
we get two (commuting) vector fields in the Lorentzian signature:
\begin{equation}
{1\over2}\gamma_2 (\phi+\ln{|\lambda|}), \qquad
{1\over2}\gamma_2 (\phi-\ln{|\lambda|}).
\end{equation}
Recalling the relation (\ref{J-gamma}) between the $\gamma$-matrices
and vector fields $J_i$, we see that what we get is exactly
the rotation VF $\xi_{rot}$ (\ref{btz-rot}) for the BTZ BH. Using
(\ref{J}) we see that the angular velocity is given by:
\begin{equation}
\Omega = {\phi\over\ln{|\lambda|}}.
\end{equation}
Recalling that for the Euclidean BH
$\phi=2\pi|r_-|/l, \ln{|\lambda|} = 2\pi r_+/l$,
we see that $\Omega=|r_-|/r_+ = J l/2 r_+^2$, which is the correct
expression for the horizon angular velocity. Here $J$ is the
BTZ BH angular momentum.

The above example shows that, at least in the BTZ case, the
angular velocity can be expressed directly in terms of the
Euclidean generator $L$. Indeed,
\begin{equation}\label{J**}
i\,\Omega= {{\rm Arccosh}\left({1\over 2}{\rm Tr}\,L \right)-
{\rm Arccosh}\left({1\over 2}{\rm Tr}\,L^\dagger \right)\over
{\rm Arccosh}\left({1\over 2}{\rm Tr}\,L \right)+
{\rm Arccosh}\left({1\over 2}{\rm Tr}\,L^\dagger \right)}.
\end{equation}
This formula is the direct analog of (\ref{J*}) in the Lorentzian
case. Let us rewrite (\ref{J**}) in a more convenient form by introducing
a function $m(L): \sigma_L {\rm Tr}\,L = m(L)^{1/2}+m(L)^{-1/2}$.
Here $\sigma_L$ is the sign of the trace. Then:
\begin{equation}\label{Omega}
\Omega = {{\rm Arg}\,m(L)
\over \ln{|m(L)|}}.
\end{equation}
The quantity $m(L)$ is called the multiplier of the transformation $L$.

So far we have only seen how the procedure (\ref{map}) works
for a group $\Sigma$ generated by a single generator, which
is the BTZ case. In the general case one has to consider
more complicated groups, with several generators. Let us
consider the case when $\Sigma$ is freely generated
by a number of loxodromic generators $L_i$. Then $\Sigma$ is called a
Schottky group. To analytically continue $\Sigma$ into a subgroup of the
$\SL(2,\R)\times\SL(2,\R)$ one has to continue each $L_i$
according to (\ref{map}) to obtain the left part of a
$\SL(2,\R)\times\SL(2,\R)$ element.
Similarly, one continues $L^\dagger$ to get the right part.
The discrete group $\Gamma$ is then freely generated by
the obtained generators.

Let us illustrate this procedure on the example of a single
asymptotic region wormhole. Let us take a Schottky group
$\Sigma$ to be generated by the following two elements:
\begin{equation}
L_1 = \exp{{i\sigma_1\over 2}(a+ib)}, \qquad
L_2 = \exp{{-i\sigma_3\over 2}(a+ib)}.
\end{equation}
Under the analytic continuation these two generators
give precisely the two VF (\ref{vf-wormhole}). The group
$\Gamma$ generated by these VF's exponentiated is the
group that is used to obtain the spacetime. The generator
of rotations of the asymptotic region is represented in the
Euclidean signature by the commutator $L=[L_1,L_2]$. It is
clear that the formula (\ref{J**}) is the analytically
continued (\ref{J*}). In other words, (\ref{J**}) goes into
(\ref{J*}) under the map (\ref{map}).

\subsection{The angular velocity as a cross-ratio}
\label{sec:cross-ratio}

In the previous subsection we have seen how Schottky groups
can be analytically continued ``back'' to obtain discrete
groups of isometries of the Lorentzian AdS${}_3$. Let us note,
however, that it only works in the opposite direction, that
is, from Euclidean to Lorentzian signature. Indeed, given the left $\gamma^L$
and right $\gamma^R$ $\SL(2,\R)$ group elements it is not clear how to
split the corresponding vector fields into the $\omega$ and $\nu$
parts, so that in the analytic continuation (\ref{map})
$\gamma^L\to L, \gamma^R\to L^\dagger$. In order to be able to
analytically continue ``forwards'' we need to understand better
how the continuation procedure works.

We have seen that non-rotating spacetimes give, in the Euclidean
signature, the groups $\Sigma$ that consists of only real
elements. As became clear in the previous subsection, the presence
of rotation is equivalent to having a complex group $\Sigma$.
In other words, one could start with a non-rotating spacetime
and the corresponding group $\Gamma\in\SL(2,\R)$. The
Euclidean group is $\Sigma=\Gamma$ in this case. Inclusion
of rotation is equivalent, in the Euclidean signature, to
a certain deformation of $\Gamma$ into a complex group
$\Sigma$. Such deformations are well-known to mathematicians:
they go under the name of quasi-Fuchsian groups, obtained
from Fuchsian ones by quasi-conformal mappings. Thus, the
idea is that, in order to construct the Euclidean group
$\Sigma$ one has to appropriately deform the Fuchsian
group $\Gamma$ of the non-rotating case. To see what
kind of deformations one has to consider we need to derive
some suggestive facts.

In this subsection we
rewrite the expression (\ref{general-J}) for the angular momentum
of an asymptotic region as a certain cross-ratio. The
idea is to consider the integral curve of $\xi_{rot}$ that
passes through the center of the diamond in which $\xi_{rot}$
is spacelike, see Fig.~\ref{fig:btz-rot}. This integral curve
can be thought of as a deformation of the real axes
due to the presence of rotation. Studying this deformation
we will understand what $\Sigma$ should be in the rotating
case. Let us first consider the BTZ case, in which the
``corners'' of the diamond are located at $\phi=0,\pi$.
Note that (\ref{J}) can be obtained as:
\begin{equation}\label{J-ratio}
\Omega=-\left({dt(\phi)\over d\phi}\right)_{\phi=\pi/2}.
\end{equation}
Here $t(\phi)$ is an equation for the integral curve of
$\xi_{rot}$ that passes through the $u=-\pi/2, v=\pi/2$ point.
To show this we write
\begin{equation}
{dt(\phi)\over d\phi} = {du+dv\over dv-du}= {1+(du/dv)\over 1-(du/dv)}.
\end{equation}
On the integral curve of $\xi_{rot}$ given by (\ref{btz-rot})
we have:
\begin{equation}
{du\over dv} = {a+b\over a-b}\, {\sin{u}\over\sin{v}}.
\end{equation}
At the point $u=-\pi/2, v=\pi/2$ or $\phi=\pi/2, t=0$
$du/dv=-(a+b)/(a-b)$ and (\ref{J-ratio}) gives $\Omega=b/a$.

\begin{figure}
\centerline{\hbox{\epsfig{figure=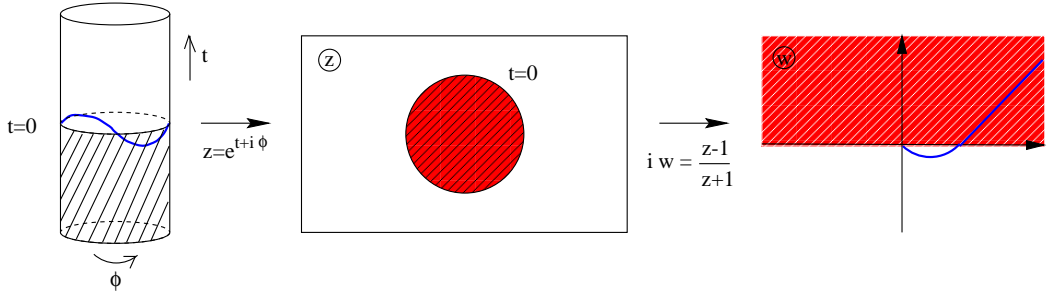,height=1.5in}}}
\caption{The (Euclideanized) boundary cylinder can be mapped into the complex
plane. The $t<0$ half of the cylinder becomes either the unit
disc or the upper half-plane. One of the integral curves of
the rotation vector field is shown. We also show its image on the
$w$-plane.}
\label{fig:maps}
\end{figure}

Let us now, using an analytic continuation, derive
a similar representation for the general expression (\ref{general-J}).
To this end, we analytically continue the boundary cylinder $\cal I$ to
a Euclidean cylinder. For this purpose we analytically continue
the (real) null coordinates $u, v$ to a new pair of complex
coordinates, which we, by abuse of notation, will also denote
by $u, v$. This is done by analytically replacing the time
coordinate $t\to i t$. We get:
\begin{equation}\label{u-v}
u = i t-\phi, \qquad v = i t+\phi
\end{equation}
The metric $ds^2=-du dv$ becomes the Euclidean metric $dt^2+d\phi^2$.
We can then conformally map the Euclidean cylinder into the
complex plane. One first maps
the $t<0$ part of the cylinder into the interior of the
unit disc, and then the later can be mapped by a fractional
linear transform into the upper half-plane, see Fig.~\ref{fig:maps}.

To represent the angular velocity as a cross-ratio we
will need an expression for the cross-ratio of 4 points on the
$z$-plane in terms of the coordinates of their pre-images on the cylinder.
We have the following straightforward identity:
\begin{equation}
z_1-z_2 = e^{t_1+i \phi_1} - e^{t_2+i \phi_2} = e^{-iu_1} - e^{-iu_2}=
e^{-{i\over2}(u_1+u_2)} 2i\, \sin{\left({u_2-u_1\over 2}\right)}.
\end{equation}
Thus, we get for the cross-ratio:
\begin{equation}
R(z_1,z_2,z_3,z_4) = {(z_3-z_1)(z_4-z_2)\over(z_3-z_2)(z_4-z_1)}=
{
\sin{\left({u_3-u_1\over 2}\right)}
\sin{\left({u_4-u_2\over 2}\right)}
\over
\sin{\left({u_3-u_2\over 2}\right)}
\sin{\left({u_4-u_1\over 2}\right)}}.
\end{equation}
Let us calculate the cross-ratio of the following four points:
\begin{equation}
z_1 = \{u, v \}, \qquad
z_2=\{u^*, v^*\} := \{ {u_p+u_{p'}\over 2}, {v_p+v_{p'}\over 2} \}, \qquad
z_3 = \{u_p, v_{p'} \}, \qquad
z_4 = \{u_{p'}, v_p \}.
\end{equation}
Here by $z=\{ u, v\}$ we mean a point $z$ on the complex $z$-plane,
whose pre-image on the cylinder has the $u, v$ complex coordinates
given by (\ref{u-v}). Thus, $z_3, z_4$ are images of the two fixed
points of $\xi_{rot}$, see Fig.~\ref{general-rot}, $z_2$ is
the ``center'' of the rectangular region in which $\xi_{rot}$
is spacelike, and $z_1$ is an arbitrary point. We get:
\begin{equation}\label{R}
R(u) = - {
\sin{\left({u-u_{p}\over 2}\right)} \over
\sin{\left({u-u_{p'}\over 2}\right)}}.
\end{equation}
Let us now show that the general formula
(\ref{general-J}) can be understood in a way similar to
(\ref{J-ratio}). It is for this purpose that we needed the
cross-ration (\ref{R}). We would like to show that the angular
velocity (\ref{general-J}) can be represented as the ratio
\begin{equation}\label{J-ratio-general}
\Omega =  \left({d\, {\rm Im}\,R(u)\over d\,{\rm Re}\,R(u)}\right)_*
\end{equation}
of the differentials of the imaginary and real part
of the cross-ratio $R(u)$ when $u$ is moved infinitesimally away from
the center $u=u^*, v=v^*$ along the integral curve of $\xi_{rot}$. This formula
is an analog of (\ref{J-ratio}).

On the integral curve of $\xi_{rot}$ we have:
\begin{equation}
{du\over dv} = - {\sin{\left({u-u_p\over2}\right)}
\sin{\left({u-u_{p'}\over2}\right)} \over
k \sin{\left({v-v_p\over2}\right)}
\sin{\left({v-v_{p'}\over2}\right)} }.
\end{equation}
Evaluating this at the center of the rectangle gives:
\begin{equation}\label{rot-1}
\left({du\over dv}\right)_* = - {\sin^2{\left({u_p-u_{p'}\over4}\right)}\over
k \sin^2{\left({v_p-v_{p'}\over4}\right)} }.
\end{equation}
Let us also find the derivative of $R(u)$ at $u=u^*$. We have:
\begin{equation}\label{rot-2}
\left({d R\over d u}\right)_* = -
{\sin{\left({u_p-u_{p'}\over2}\right)} \over
2 \sin^2{\left({u_p-u_{p'}\over4}\right)} }.
\end{equation}
We can now express the differentials of the real and imaginary parts
of the cross-ratio as:
\begin{equation}\label{rot-3}
d\,{\rm Re} 2 R(u,v) = \left({d R\over d u}\right)\,du+
\left({d \bar{R}\over d v}\right)\,dv, \qquad
d\,i\,{\rm Im} 2 R(u,v) = \left({d R\over d u}\right)\,du-
\left({d \bar{R}\over d v}\right)\,dv.
\end{equation}
Combining (\ref{rot-1}), (\ref{rot-2}) and (\ref{rot-3}), we
see that (\ref{J-ratio-general}) indeed equals to (\ref{general-J}).
This means, in particular, that $\Omega=0$ whenever the image of
the integral curve of $\xi_{rot}$ on the $z$-plane is circular.
In this case the cross-ratio $R(u,v)$ is real (equal to one), and
(\ref{J-ratio-general}) is zero.

\subsection{The Fenchel-Nielsen deformation}

In the previous subsection we have seen that the image of the
integral curve of $\xi_{rot}$ under the analytic continuation
of the boundary cylinder allows one to determine the
angular velocity, see formula (\ref{J-ratio-general}). Let us
now explicitly find the image of this curve. Thus, consider
the integral curve of $\xi_{rot}$ that passes through the center of
the diamond in which $\xi_{rot}$ is
spacelike. Let us first consider the case relevant for BTZ BH, in
which the fixed points of $\xi_{rot}$ are at $\phi=0,\pi, t=0$,
see Fig.~\ref{fig:maps} for a drawing of the integral curve.
The rotation VF is given by (\ref{btz-rot}).
It is not hard to find the integral curve of interest. One gets:
\begin{equation}\label{curve}
\ln{\left(-\tan{u/2}\right)} = {a+b\over a-b}\ln{\left(\tan{v/2}\right)}.
\end{equation}
The minus sign on the left hand side is necessary to make the argument
of the logarithm positive. When $b=0$ (no rotation) this equation implies
$-u=v$, which is the equation of the real axes. For $b\not= 0$ the arguments
of both logarithms at $u=-\pi/2, v=\pi/2$ are one, so that the equation
is satisfied. Thus, the curve (\ref{curve}) passes through the
center of the diamond $u=-\pi/2, v=\pi/2$ for any $b$. Let us
now find its image on the $w$-plane. As in subsection
(\ref{sec:cross-ratio}) we replace $t$ by $it$ everywhere, so that
$u, v$ coordinates become those given by (\ref{u-v}). Using the
fact that $z=e^{-i u}$ we see that:
\begin{equation}
-\tan{u/2} = - {e^{iu/2} - e^{-iu/2}\over i(e^{iu/2}+e^{-iu/2})} =
{1\over i}\, {z-1\over z+1} = w.
\end{equation}
Similarly,
\begin{equation}
\tan{v/2} = \bar{w}.
\end{equation}
Thus, the integral curve (\ref{curve}) goes on the $w$-plane
into the curve:
\begin{equation}\label{image}
w^{1+i\, \Omega} = \bar{w}^{1-i\, \Omega}.
\end{equation}
Here $\Omega=i b/a$ is the analytically continued angular velocity. We
shall denote it by the same letter, hoping that this will not
lead to any confusion. The equation (\ref{image}) can be understood as
the image $f^\Omega = \overline{f^\Omega}$ of the real line $w=\bar{w}$
under the conformal map
\begin{equation}\label{f-J}
f^\Omega(w) = w^{1+i\, \Omega}.
\end{equation}
The angular velocity can be expressed as:
\begin{equation}
\Omega = \left({d\, {\rm Im}\, f^J\over dw}\right)_{w=1}.
\end{equation}
This is clearly the same formula as (\ref{J-ratio}), but
in the Euclidean domain. Similarly, one can integrate
the general VF (\ref{general-rot}). As is not hard to show,
the relevant integral curve is now given by:
\begin{equation}
\ln{\left(-{
\sin{\left({u-u_{p}\over 2}\right)} \over
\sin{\left({u-u_{p'}\over 2}\right)}}\right)} = {1+\Omega\over 1-\Omega}
\ln{\left(-{
\sin{\left({v-v_{p}\over 2}\right)} \over
\sin{\left({v-v_{p'}\over 2}\right)}}\right)}
\end{equation}
Or, using the cross-ration (\ref{R}) one can rewrite this equation as:
\begin{equation}
\ln{R(u)} = {1+\Omega\over 1-\Omega} \ln{R(v)}.
\end{equation}
One can now find the image of this curve on the $w$-plane. It is
clear that we get an equation similar to (\ref{image}), with the
cross-ratio $R$ instead of $w$. Since 3 points on the
$w$-plane can always be mapped to $0,1,\infty$, we can always
reduce the general case to that described by (\ref{image}) by
a conformal transformation. Thus, we will only consider the
curve (\ref{image}) below.

Let us now interpret all these facts. The integral curve
of $\xi_{rot}$ can be thought of as the deformation of
the $t=0$ axes on the boundary cylinder due to rotation.
Then on the Euclidean plane this deformation is described
by the map $f^\Omega$ (\ref{f-J}). This is exactly the map
that appears in the quasi-Fuchsian deformations due to
the Fenchel-Nielsen twist. We now turn to review some
necessary facts about the later. Our main reference on this
is \cite{Wol-1}, see also \cite{Wol-2}.

\begin{figure}
\centerline{\hbox{\epsfig{figure=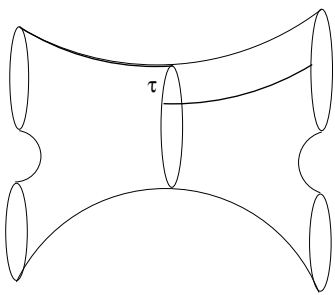,height=1.5in}}}
\caption{Any Riemann surface can be glued out of 3-holed spheres. The
glueing prescription involves a measure of how much two boundary
geodesics are rotated with respect to each other: the twist $\tau$.}
\label{fig:twist}
\end{figure}

The Fenchel-Nielsen twist was introduced as a real modulus for
Riemann surfaces. The idea is that any surface can be glued out of
3-holed spheres, sometimes called trinions. Each trinion is
characterized by the three length of the boundary geodesics.
Glueing a pair of boundaries of two trinions (or two boundaries of
one and the same trinion) one can make a twist, called the
Fenchel-Nielsen twist. The twist is measured as the geodesic
distance between the open ends of two geodesics incident on the
geodesic of the twist, see Fig.~\ref{fig:twist}. Let us see how
these twists give real coordinates on the moduli space. To think
of a Riemann surface of genus $g$ as composed out of trinions one
has to introduce a maximal set of non-intersecting, homotopy
non-trivial curves. The number of curves in such a set is $3g-3$.
They can be chosen, however, in many different ways. Cutting the
surface along these curves one obtains $2g-2$ trinions. The length
of these geodesic curves together with the Fenchel-Nielsen twists
make $6g-6$ real numbers, which serve as a system of coordinates
on the moduli space. Given a set of trinions together with a
glueing prescription (the twists) there is an effective procedure
of obtaining the resulting Riemann surface: one finds its Fuchsian
group by combining the Fuchsian groups of the individual trinions,
see \cite{Abikoff,Wol-1,Maskit}.

In \cite{Wol-1} the quasi-conformal map associated with the
Fenchel-Nielsen twist was studied. Recall, that the Teichmuller
space $T_g$ can be realized as the space of all Fuchsian groups
$\Gamma^\mu$ obtained from a base point group $\Gamma$ by
quasi-conformal deformations $\Gamma^\mu = f^\mu\circ\Gamma\circ
{f^\mu}^{-1}$. Here $f^\mu$ is a solution of the Beltrami equation
$f^\mu_{\bar{z}} = \mu f^\mu_z$. To obtain Fuchsian groups the
Beltrami differential should satisfy a ``reality condition'':
$\mu(z,\bar{z})=\overline{\mu(\bar{z},z)}$. This model for $T_g$
is sometimes inconvenient, for the dependence of $f^\mu$ on $\mu$
is only real analytic, not complex analytic. Thus, the complex
structure on $T_g$ is not transparent in this model. For this
reason is is often preferable to use another model, in which
$\mu=0$ in the lower half-plane $\bar{\bf H}$. The corresponding
quasi-conformal maps $f^\mu$ do not leave the real axes invariant.
Instead, it is mapped into a Jordan curve, and the upper
half-plane is mapped into the interior of this curve. This is
exactly what we need, for, as we saw, in the deformation due to
rotation the real axes is mapped into a certain curve, see, e.g.,
Fig.~\ref{fig:wormhole-rot}. The corresponding groups $\Gamma^\mu$
are called quasi-Fuchsian. They are subgroups of $\SL(2,\C)$,
which is again what we expect to be relevant in our case. For more
information on Teichmuller spaces and quasi-Fuchsian groups we
send the reader to \cite{Book}.

In \cite{Wol-1} the author studied the quasi-Fuchsian realization
of the Fenchel-Nielsen twist. As is explained in \cite{Wol-1}, the
Beltrami differential corresponding to an infinitesimal twist on
the geodesic $0,\infty$ (and the case of a cyclic group $\Gamma$)
is given by
\begin{equation}
\mu(w,\bar{w}) = {i \epsilon \bar{w}\over 2 w} \chi({\rm Arg}\, w),
\qquad  \chi({\rm Arg}\, w) = (\theta_2-\theta_1)^{-1}
\chi_{[\theta_1,\theta_2]}({\rm Arg}\, w).
\end{equation}
Here $\chi_{[\theta_1,\theta_2]}({\rm Arg}\, w)$ is the characteristic function
of the interval $[\theta_1,\theta_2]\subset[0,2\pi]$. We switched to
calling our complex variable $w$ because we now use the upper half-plane
model for $\bf H$, see Fig.~\ref{fig:maps}. The first
variation of $f^\mu$ is found \cite{Wol-1} to be:
\begin{equation}\label{dot-f}
\dot{f}(w) = w\left( \int_0^{{\rm Arg}\, w} \chi(t) dt + {i\over 2\pi} \ln{w}\right).
\end{equation}
Let us show that the twist parameter $\epsilon$ can be
represented by a formula suggestive of (\ref{J-ratio}),
(\ref{J-ratio-general}).
Let us consider the image of the real axes under the map
$f^\mu$. We have:
\begin{equation}\label{f-first}
f^\mu(x) = x+ {i \epsilon\over 2\pi} x\,\ln{x}.
\end{equation}
Note that this is a curve that passes intersects the real axes at
point $x=1$; this curve is similar to the integral curve of the
rotation VF $\xi_{rot}$, see Fig.~\ref{fig:maps}(c). The
deformation parameter $\epsilon$ can now be represented as:
\begin{equation}\label{twist}
{\epsilon\over 2\pi} = \left({d\, {\rm Im}\, f^\mu(x)\over d x}\right)_{x=1}
\end{equation}
Notice, moreover, that the conformal part (\ref{f-first}) of the described
quasi-conformal map $f^\mu$ coincides with the map $f^\Omega$
(\ref{f-J}), at least for small $\Omega$ when we can compare $f^\Omega$
to the known first variation of $f^\mu$. Thus, at least for a
cyclic group $\Gamma$, and infinitesimally, the real axes
gets deformed due to rotation exactly like in the Fenchel-Nielsen
twist. This suggests that the twist parameter of the Euclidean surface
should be thought of as a measure of the angular velocity in the
corresponding Lorentzian BH.

Let us extend this interpretation to the case of an arbitrary, not
necessarily cyclic Fuchsian group $\Gamma$. This requires some
background. As is explained in \cite{Wol-1}, the corresponding
quasi-conformal map $f^\mu$ is constructed by superposing maps
(\ref{dot-f}) for all images of the axes of the twist. Let us
explain this in more detail. Consider a Riemann surface $X$ with
boundary, and uniformize $X$ by a Fuchsian group $\Gamma$, which
consist of only hyperbolic elements. One should think of $X$ as
being the $t=0$ plane with all the asymptotic regions detached,
see Fig.~\ref{fig:disconnect}. Then boundary geodesics are in
one-to-one correspondence with asymptotic regions. Among the
elements of $\Gamma$ there are some whose axes project to boundary
geodesics. Let us consider one of the boundary geodesics, and
denote it by $\alpha$. Consider elements whose axes on $\bf H$
project to $\alpha$ on $X$. Such axes in $\bf H$ are in one-to-one
correspondence with certain conjugacy classes in $\Gamma$. For
every such axe there exists a hyperbolic cyclic subgroup $<A>$
generated by an element $A\in\Gamma$, such that $|{\rm Tr}\, A| =
2\cosh{(l/2)}$, where $l$ is the length of the boundary, and the
axe of $a$ projects to the boundary geodesic. Other axes that
project to $\alpha$ in $X$ are in one-to-one correspondence with
conjugacy classes $\gamma A \gamma^{-1}, \forall\gamma\in\Gamma$.
The axe of $<A>$ separates $\bf H$ into two half-planes. Let us
call the {\it boundary half-plane} the one that is precisely
invariant under $<A>$ in $\Gamma$. The other half-plane, which is
not precisely invariant (unless $\Gamma$ is elementary, as in the
case of BTZ BH), is called the {\it action half-plane}. One should
think of the boundary half plane of $<A>$ as one of the copies of
the covering space of the asymptotic region that is attached to
$\alpha$. Other copies of this covering space are obtained as
boundary half-planes of the conjugate axes $\gamma <A>
\gamma^{-1}$.

Let us now consider the effect of a Fenchel-Nielsen twist on
$\alpha$. In the quasi-Fuchsian model a twist is described by a
quasi-conformal map $f^\mu$ that maps the real axes into a Jordan
curve. Let us consider the effect of $f^\mu$ on the boundary
half-plane of $<A>$. The axe of $<A>$ can always be mapped by an
$\SL(2,\R)$ transformation to the geodesic $0,\infty$. The
boundary half-plane is then the first quadrant. In the
neighborhood of the positive real axes the map $f^\mu$ becomes
conformal. It is, however, no longer true that this conformal map
is given by (\ref{f-J}). As is shown in \cite{Wol-1}, for an
infinitesimal twist, this map is given by an infinite
superposition of such maps from all conjugacy classes of $<A>$.
Thus, it will no longer be true that the image (\ref{image}) of
the integral curve of $\xi_{rot}$ on the $w$-plane is given by the
image of the real axes (its part in the corresponding boundary
half-plane) under the Fenchel-Nielsen quasi-conformal deformation.
This correspondence only holds for a cyclic group $\Gamma$, that
is for the BTZ case. However, certain other relation is still
valid even in the general case. Recall that the angular velocity
$\Omega$ was expressed (\ref{Omega}) as the ratio of the imaginary
and real parts of the muliplier of the transformation $L$
generating the asymptotic region. The above discussion suggests
that we have to identify $L$ with the deformation of $A$ due to a
twist. Let us study consequences of such an identification.

Thus, let us set $L=A^\tau$, where $A^\tau$ is the quasi-Fuchsian
deformation of $A$. We have $\ln{|m(A^\tau)|} = l$, so to find the
corresponding angular velocity we only need to find the argument
of the multipler $m(A^\tau)$. Wolpert \cite{Wol-1} proves the
following fact (Theorem 4.4)
\begin{equation}\label{fact}
{\cal P}(\Theta_A; B) = 2 \pi i \sigma_{B} \sinh{(l/2)}\,
{\partial \ln{m(\alpha)}\over \partial \tau(\beta)}
\end{equation}
Here $\sigma_{B}$ is the sign of ${\rm Tr}\,B$, and ${\cal
P}(\Theta_A; B)$ is the period of a quadratic differential
$\Theta_A$ along the geodesic $\beta$. The period of a quadratic
differential $\varphi$ along a geodesic $\alpha$ corresponding to
$A$ is defined as:
\begin{equation}
{\cal P}(\varphi, A)= {1\over 2} \int_w^{Aw} \varphi (cw^2 +
(d-a)w-b) dw.
\end{equation}
The quadratic differential $\Theta_A$ is the Poincare series
constructed using the element $A$. Let us give a definition. For
\begin{eqnarray*}
A = \left(\begin{array}{cc} a & b \\ c & d \end{array}\right)
\end{eqnarray*}
define
\begin{equation}
\omega_A = ({\rm Tr}^2 A-4)(cw^2 +(d-a)w - b)^{-2}.
\end{equation}
Then
\begin{equation}\label{Poincare}
\Theta_A = \sum_{B\in <A>\backslash\Gamma} \omega_{B^{-1} A B}.
\end{equation}
Now, the result (\ref{fact}) can be understood by recalling that,
as was shown by Hejhal \cite{Hejhal}, the period ${\cal P}(\phi;
A)$ is the first variation of the trace of $A$ under a
quasi-conformal map whose Schwarzian derivative is $\phi$. Wolpert
\cite{Wol-1} proves that $\Theta_A$ is exactly the Schwarzian
derivative of the quasi-conformal map corresponding to an
infinitesimal Fenchel-Nielsen twist:
\begin{eqnarray}
\tau \Theta_A(w) = {\cal S}(f^\tau; w).
\end{eqnarray}
Here $f^\tau$ is the conformal part of the quasi-conformal map,
that is, the restriction of $f^\tau$ to the lower half-plane.
Combining results of Hejhal and Wolpert one gets (\ref{fact}).
Setting $B=A$ we see that, for an infinitesimal twist, $\rm Arg$
of $m(A^\tau)$ can be expressed through the real part of the
period. Recalling (\ref{Omega}) we get:
\begin{eqnarray}\label{Omega-twist}
\Omega = - {\sigma_A\over 2\pi l \sinh{(l/2)}} {\cal
P}(\tau\,\Theta_A; A).
\end{eqnarray}
In case when $\Gamma$ is cyclic $\Gamma=<A>$, the period can be
easily computed. One gets ${\cal P}(\Theta_A; A) = - \sigma_A l
\sinh{(l/2)}$. This means that for a cyclic group, like the one of
the BTZ BH, the angular velocity is simply the Fenchel-Nielsen
twist $\Omega=\tau/2\pi$. For a general group the period does not
seem to be computable in any explicit way, and the relation
between the angular velocity and the twist parameter becomes more
complicated (\ref{Omega-twist}). It still, however, admits a
simple interpretation. Indeed, since $\tau \Theta_A$ is the
Schwarzian derivative of the twist map $f^\tau$, it can be
interpreted as the (holomorphic part of the) stress-energy tensor
of a CFT. Then (\ref{Omega-twist}) is just the higher genus
version of the standard CFT formula:
\begin{eqnarray}
\Delta = {1\over 2\pi i} \int T(z) z dz.
\end{eqnarray}
Here $\Delta$ is the conformal dimension, which is related to the
angular momentum as $J=\Delta-\bar{\Delta}$, and the angular
momentum is $J=2\Omega r_+^2, l=2\pi r_+$. Our expression for the
angular velocity (\ref{Omega-twist}) is consistent with this CFT
interpretation.

With the result (\ref{Omega-twist}) at hand we are finally ready
to formulate the prescription for the analytic continuation.
Consider a rotating black hole spacetime. First, set the angular
momentum of all asymptotic regions to zero. In this case, there
exists a plane of time symmetry. Its geometry is that of ${\bf
H}/\Gamma$, for some Fuchs group $\Gamma$. The corresponding
Euclidean space is $M=\H/\Gamma$, where now $\Gamma$ is viewed as
a subgroup of $\SL(2,\C)$ acting on $\H$ by isometries. The
boundary of $M$ is the Schottky double $\tilde{X}={\cal C}/\Gamma$
of ${\bf H}/\Gamma$, where $\cal C$ is the complement of the set
of fixed points of $\Gamma$ on $\C$. Let us now turn on the
rotation. On the Euclidean side this is achieved by a
quasi-conformal map, mapping the real axes into a Jordan curve.
The quasi-conformal map is the one that corresponds to the
Fenchel-Nielsen deformation. Under the quasi-conformal map the
Fuchsian group $\Gamma$ becomes a quasi-Fuchsian group
$\Sigma=\Gamma^\tau\subset\SL(2,\C)$. The analytic continuation of
our rotating spacetime is given by $M=\H/\Gamma^\tau$. Its
boundary is given by $\tilde{X} ={\cal C}^\tau/\Gamma^\tau$, where
${\cal C}^\tau$ is the complement of the set of fixed points of
$\Gamma^\tau$ in $\C$. The surface $\tilde{X}$ is of the same
genus as in the non-rotting case. Its moduli space is in
one-to-one correspondence with the parameter space of our
spacetime. In particular, the horizon size is given by the length
of a particular geodesic on $\tilde{X}$, and the horizon angular
velocity is given by the period (\ref{Omega-twist}) of a certain
quadratic differential (Poincare series) along the geodesic. The
expression (\ref{Omega-twist}) is valid for infinitesimal twists.
For a finite twist one has to use the expression (\ref{Omega}).
Finally, let us note that $\tilde{X}$ can be thought of as the
Schottky double of $X$ even in the rotating case. In this case the
anti-holomorphic map $\sigma: X=\tilde{X}/\sigma$ is given by the
inversion in the Jordan curve.

\section{Discussion}
\label{sec:disc}

Thus, our analytic continuation procedure is that of continuing
the discrete groups. One has to start from a non-rotating BH, with
its discrete group being a subgroup of $\SL(2,\R)$ and the
analytic continuation being trivial. One should then turn on the
rotation, which deforms the discrete group $\Gamma$ into a
subgroup $\Gamma^\tau$ of $\SL(2,\C)$. The angular velocity of an
asymptotic region is given by a certain period, see
(\ref{Omega-twist}). We established this identification for
infinitesimal twists. For a finite deformation the angular
velocity of an asymptotic region is given by (\ref{Omega}). For a
cyclic group $\Gamma$ the angular velocity is essentially the
twist $\Omega=\tau/2\pi$. It could be that this also holds for a
general group, which would be the case if ${\cal P}(\Theta_A; A) =
-\sigma_A l \sinh{(l/2)}$ for a general $\Gamma$. We were not able
to establish or disprove this. If this is not true, the best one
has for a finite twist would be the formula (\ref{Omega}).

Our construction solves the problem of classification of rotating
wormhole spacetimes. Indeed, similarly to the non-rotating case,
in which BH's are classified by their $t=0$ plane geometry, the
rotating spacetimes are classified by the conformal geometry of
the boundary of the corresponding Euclidean spaces. Moduli of this
Eucildean boundary are in correspondence with the BH parameters.
In particular, the horizon angular velocities are given by
(\ref{Omega}), where $L$ are group generators that project to the
horizon geodesics.

Let us conclude with a few comments as to the thermodynamics of
black and wormhole spacetimes. Given a spacetime one can construct
the corresponding Euclidean manifold using our analytic
continuation procedure. The manifold $M$ is a handlebody. One can
then evaluate the partition function, which is the value of the
Euclidean Einstein-Hilbert action on $M$. This problem was studied
in \cite{Riemann}. The calculation involves a particular
regularization procedure. Ofter this is done, the value of the
partition function is show to be given by the Liouville action
evaluated on the canonical (Poincare) Liouville field, see
\cite{Riemann}. Thus, the partition function $Z$ is a particular
function of the moduli of the Euclidean boundary $\tilde{X}$. As
was shown in \cite{Takht}, $Z$ is the Kahler potential for the
Weyl-Peterson symplectic structure on the Teichmuller space $T_g$.
Then one can take the length $l$ and the angular momentum $\Omega$
(\ref{Omega}) of the geodesics corresponding to asymptotic regions
as part of the moduli. One can then form out of $l, \Omega$ the
intensive thermodynamic quantities: the inverse temperature
$\beta$ and the parameter $\Phi$ conjugate to the angular momentum
$J$ . Then the derivative of $-\ln{Z}$ with respect to $\beta$ and
$\Phi$ should give the mass $M$ and the angular momentum $J$
correspondingly. This is a non-trivial differential equation on
$Z$, and thus on the Liouville action. Then, performing the
Legendre transform to the variables $M, J$ for every asymptotic
region, one should find that the result of the transform (entropy)
is given by the sum of length $l$ of all boundary geodesics of
$X$. This is another very non-trivial equation $Z$ has to satisfy.
Thus, the black hole interpretation of Riemann surfaces leads us
to conjecture some rather non-trivial identities for the Kahler
potential $\ln{Z}$ on the Teichmuller space $T_g$. It would be of
considerable interest to study this structure in more details. We
hope to return to this problem in the future.

\noindent
{\large \bf Acknowledgments}

I would like to thank L.\ Friedel, J.\ Hartle and G.\ Horowitz
for discussions. I am grateful
to S.\ Wolpert for correspondence. The author was supported by
the NSF grant PHY00-70895.

\eject

\appendix
\section{Some properties of the groups of isometries}
\label{app}

\bigskip
\noindent{\bf A. Lorentzian case}
\bigskip

Here we review some useful facts about the group of isometries of
the Lorentzian AdS${}_3$. This material is widely known, see,
e.g., \cite{Rot}.

Let us recall that the Lorentian AdS${}_3$ can be defined as a
quadric $-U^2-V^2+X^2+Y^2=-l^2$ in $\R^{2,2}$. The metric is
given by: $ds^2=-dU^2-dV^2+dX^2+dY^2$. Thus, its group
of isometries is ${\rm O}(2,2)$. It is often more convenient to
introduce another set of coordinates $t,\rho,\theta$ defined by:
\begin{eqnarray*}\label{t-rho-theta}
U={1+\rho^2\over 1-\rho^2}\cos{t}, &\qquad&
V={1+\rho^2\over 1-\rho^2}\sin{t} \\
X={2\rho\over 1-\rho^2}\cos{\theta}, &\qquad&
Y={2\rho\over 1-\rho^2}\sin{\theta}.
\end{eqnarray*}
The metric then takes the following simple form:
\begin{equation}
ds^2 = - \left( {1+\rho^2\over 1-\rho^2} \right)^2 dt^2 +
\left( {2\over 1-\rho^2} \right)^2 (d\rho^2+\rho^2d\theta^2).
\end{equation}
Note that the constant $t$ planes in this model are all isometric to
the hyperbolic plane (in the Poincare unit disc model). The
conformal infinity $\cal I$ is the (timelike) unit cylidner $t,\theta$.

Another very convenient model, the one best suited for calculations,
is that of the $\SL(2,\R)$ group manifold, or, more precisely,
its universal cover. Note that the equation of the quadric
defining AdS${}_3$ can be rewritten as a requirement that
the following $2\times 2$ matrix has the unit determinant:
\begin{equation}
{\bf x} = \left( \begin{array}{cc}
  U+X & Y+V \\
  Y-V & U-X \end{array} \right)
\end{equation}
This makes it clear that AdS${}_3$ can be realized as the universal
cover of the $\SL(2,\R)$ group manifold. In this model the
metric is just the natural metric on the group manifold:
\begin{equation}
ds^2 = {1\over2}{\rm Tr}({\bf x}^{-1} d{\bf x} {\bf x}^{-1} d{\bf x}).
\end{equation}
This model also makes it clear that isometries can
be realized as the left and right action of $\SL(2,\R)$, that
is ${\bf x} \to g {\bf x} \tilde{g}^{-1}$. More precisely, the
connected subgroup of the group of isometries is:
\begin{equation}\label{lor-isom}
{\rm SO}_0(2,2) = {\rm SL}(2,\R)\otimes {\rm SL}(2,\R)/\Z_2,
\end{equation}
where $\Z_2$ is generated by the element $\{-1,-1\}$. One has to
mod out by the action of $\Z_2$ since the transformation with
$-g,-\tilde{g}$ generates the same isometry.
The full group of isometries contains two additional generators. They
are given by the following reflections:
\begin{eqnarray}\label{refl}
\pi: (X,Y,U,V) \to (X,-Y,U,-V), \\ \nonumber
\Pi: (X,Y,U,V) \to (X,Y,U,-V).
\end{eqnarray}

Let us also give a parametrization of the $\SL(2,\R)$ group
manifold in terms of coordinates the $t,\rho,\theta$ introduced
in (\ref{t-rho-theta}). We have:
\begin{equation}\label{point}
{\bf x} = {1+\rho^2\over 1-\rho^2} {\bf \omega}(t) +
{2\rho\over 1-\rho^2} {\bf \gamma(\theta)},
\end{equation}
where
\begin{equation}
{\bf \omega}(\alpha) = \cos(\alpha) {\bf 1} + \sin(\alpha){\bf \gamma}_0,
\qquad
{\bf \gamma}(\alpha) = \sin(\alpha) {\bf \gamma}_1 +
\cos(\alpha){\bf \gamma}_2,
\end{equation}
and $\gamma_a$ are the $\gamma$-matrices in 2+1 dimensions:
\begin{eqnarray}\label{gamma-matrices}
\gamma_0 = \left(\begin{array}{cc}
0 & 1 \\ -1 & 0 \end{array}\right), \qquad
\gamma_1 = \left(\begin{array}{cc}
0 & 1 \\ 1 & 0 \end{array}\right), \qquad
\gamma_2 = \left(\begin{array}{cc}
1 & 0 \\ 0 & -1 \end{array}\right).
\end{eqnarray}
These matrices satisfy:
\begin{equation}
{\bf \gamma}_a {\bf \gamma}_b = \eta_{ab} {\bf 1} - \varepsilon_{ab}^c\gamma_c,
\end{equation}
where $a,b = 0,1,2$, $\eta_{ab}={\rm diag}(-1,1,1)$ is the
three-dimensional Minkowski metric which is used to raise and
lower indices, and $\varepsilon^{abc}$ is the Levi-Civita symbol
with $\varepsilon^{012}=1$.

Let us now review some facts about the vector fields in AdS${}_3$.
Vector fields in this spacetime are in one-to-one correspondence
with elements of the Lie algebra corresponding to the group
of isometries. As we saw, the later can be either realized as ${\rm SO}(2,2)$,
or as $\SL(2,\R)\times\SL(2,\R)$. The first realization of the
isometry group corresponds to the quadric model. In this
model the VF are given by $J_{XY}=X\partial_Y-Y\partial_X,
J_{YV}=Y\partial_V+V\partial_Y$, etc. In the group manifold
model the VF are the left and right invariant
vector fields on the group. They can be related to the VF of the
quadric model as follows:
\begin{eqnarray*}
J_1 = -{1\over2}(J_{XU}+J_{YV})=\sin{u}\partial_u, &\qquad&
\tilde{J}_1 = -{1\over2}(J_{XU}-J_{YV})=\sin{v}\partial_v, \\
J_2 = -{1\over2}(J_{XV}-J_{YU})=-\cos{u}\partial_u, &\qquad&
\tilde{J}_2 = -{1\over2}(J_{XV}+J_{YU})=-\cos{v}\partial_v, \\
J_3 = -{1\over2}(J_{XY}-J_{UV})=\partial_u, &\qquad&
\tilde{J}_2 = {1\over2}(J_{XY}+J_{UV})=\partial_v.
\end{eqnarray*}
This table is from \cite{Rot}. The VF $J_i, \tilde{J}_i$
are generators of the two copies of the Lie algebra ${\mathfrak sl}(2)$.
Here we have also indicated what these VF
become on the conformal boundary cylinder $\cal I$: $u=t-\phi, v=t+\phi$
are the usual null coordinates on $\cal I$. Let us also
note that the generators $J_i$ can be expressed in terms of the
$\gamma$-matrices. We have:
\begin{eqnarray}\label{J-gamma}
J_1 = -{1\over 2}\gamma_1, \qquad
J_2 = -{1\over 2}\gamma_2, \qquad
J_3 = -{1\over 2}\gamma_0.
\end{eqnarray}

Note that not all of $\SL(2,\R)$ group manifold can be reached
from the identity by the exponential map. Since the group ${\rm
SL}(2,\R)$ is AdS space, we can draw for it a Penrose diagram, see
\cite{Rot}. We get an infinite strip. It is then easy to see that
only the elements within the light cone built on the identity
element (the origin of AdS) are accessible from the origin by the
exponential map. The elements outside the light cone, which cannot
be reached, satisfy ${\rm Tr}(g) < -2$.

As an example of the exponential map, let us work out the equation for
geodesics passing through the origin of AdS. Such geodesics correspond
to one-parameter subgroups of the type $e^{s{\bf n}}$, where
$s\in\R, {\bf n}\in{\mathfrak sl}(2)$. Let us take ${\bf n} =
{\bf \gamma}_0 + \kappa {\bf \gamma}(\theta)$, where $\kappa\in\R$.
We then get:
\begin{equation}\label{geod}
\gamma = e^{s{\bf n}} =
{\rm cs}(s) + {\rm sn}(s)(\gamma_0 + \kappa\gamma(\theta)),
\end{equation}
where
\begin{equation}\label{cs}
{\rm cs}(s) = \cosh{\left(s\sqrt{\kappa^2-1}\right)}, \qquad
{\rm sn}(s) = (\kappa^2-1)^{-1/2}\sinh{\left(s\sqrt{\kappa^2-1}\right)}.
\end{equation}

\bigskip
\noindent{\bf B. Euclidean case}
\bigskip

Here we review some necessary info about the Euclidean AdS${}_3$,
which is the same as the 3D hyperbolic space. Our main references
here are books \cite{Hyperb,Fenchel}

The 3D hyperbolic space $\H$ can be realized as a quadric
$-U^2+V^2+X^2+Y^2=-l^2$ in Minkowski space $\R^{1,3}$. The
metric is $ds^2=-dU^2+dV^2+dX^2+dY^2$. Thus, its group
of isometries is the Lorentz group ${\rm SO}(1,3)$. The hyperbolic space
can be conveniently visualized as one of the sheets of the
two sheeted hyperboloid in $\R^{1,3}$. Being a homogeneous
${\rm SO}(1,3)$ manifold, the hyperbolic space can also
be thought of as a coset ${\rm SO}(1,3)/{\rm SO}(3)$. Here
${\rm SO}(3)$ is the stabilizer of a point in $\H$. Since
Lorentz group is isomorphic to $\SL(2,\C)$, this coset can
also be realized as $\SL(2,\C)/{\rm SU}(2)$. The later has
the canonical representation in terms of $2\times 2$ unitary
unimodular matrices:
\begin{equation}
{\bf x} = \left( \begin{array}{cc}
  U+V & X+iY \\
  X-iY & U-V \end{array} \right)
\end{equation}
The condition of ${\rm det}\,{\bf x}=1$ is exactly the equation
of the defining quadric. In this representation the action of
isometry group $\SL(2,\C)$ is realized as the vector
representation: ${\bf x}\to {\bf S}{\bf x}{\bf S}^\dagger$.

A model most convenient for our purposes is that of the upper half-space.
In this model $\H$ is just the upper half of $\R^3$ with
the metric:
\begin{equation}
ds^2={1\over\xi^2}\left(d\xi^2+dx^2+dy^2\right).
\end{equation}
One can think of this space as the space of (positive) imaginary
quaternions, see \cite{Fenchel}. Then the action of the isometry
group $\SL(2,\C)$ is realized as fractional linear transformations
on quaternions. One then sees that the element $-1\in{\rm SL}(2,\C)$
does not act, so that the group of isometries of $\H$
is actually ${\rm PSL}(2,\C)={\rm SL}(2,\C)/\Z_2$.

A very convenient description of elements of the isometry group is
in terms of fixed points of its action on the boundary of $\H$.
The boundary is just the (compactified) complex plane. Recall that
elements of ${\rm SL}(2,\C)$ can be divided into conjugacy
classes, according to the value of the trace of the corresponding
matrix. Thus, $L\in{\rm SL}(2,\C)$ is called {\it elliptic} if
$0\leq{\rm Tr}^2(L)<4$, {\it parabolic} if $0\leq{\rm Tr}^2(L)=4$,
{\it hyperbolic} if $0\leq{\rm Tr}^2(L)>4$, and {\it loxodromic}
if $0\leq{\rm Tr}^2(L)\in\C\backslash [0,4]$. Any transformation
$L$ has two fixed points (not necessarily distinct). Elliptic
elements have two fixed points, one located in the upper half of
$\C$ and the other being its mirror image with respect to the real
axes. Parabolic elements have their fixed points coinciding, and
located on the (extended) real axes. Hyperbolic elements have two
distinct fixed points located on the (extended) real axes.

Any transformation can be parametrized by positions of its fixed points
and the value of the so-called multiplier. The canonical representation
is:
\begin{equation}\label{canonical}
{L w - b\over L w - a} = m {w - b\over w - a},
\qquad |m| > 1.
\end{equation}
Here $a,b$ are the fixed points, $m$ is the multiplier of the
transformation, and $w$ is the coordinate on the complex plane. The
matrix corresponding to such a transformation is given by:
\begin{equation}\label{L}
L = {\sqrt{m}\over a-b} \left( \begin{array}{cc}
a - {b\over m} & ab({1\over m}-1) \\
1-{1\over m} & {a\over m} - b \end{array} \right)
\end{equation}
with an appropriate convention as to what branch of the square root
should be taken. Geometrically, the transformation given by (\ref{L})
is a shift in the amount of $\ln{|m|}$
in the direction of the geodesics going between the
fixed points, plus a rotation by an angle ${\rm Arg}(m)$.

The group ${\rm SL}(2,\C)$ can be represented as a direct product of two
copies of the (complexified) ${\rm SU}(2)$. This is the standard construction
of the vector representation of the Lorentz group. It proceeds as follows.
Any Lorentz transformation can be represented as a combination of a boost and
rotation. This is done by representing the transformation as:
\begin{equation}
L = e^{{i\over2}\sigma_i (\omega^i-i\nu^i)},
\end{equation}
where $K_i=-i\sigma_i/2$ are the generators of boosts, and $J_i=\sigma_i/2$ are
the generators of rotations. One can introduce the following (complex) linear
combinations of these generators:
\begin{equation}
N_i = J_i+iK_i, \qquad \tilde{N}_i = J_i - iK_i,
\end{equation}
in terms of which the transformation becomes
\begin{equation}
L = e^{{i\over2}N_i ( \omega^i - i\nu^i) +
{i\over2}\tilde{N}_i (\omega^i+i\nu^i)}.
\end{equation}
The two sets of generators $N_i,\tilde{N}_i$ each form the Lie
algebra ${\mathfrak su}(2)$, and the above transformation is just
a composition of the two (complexified) ${\rm SU}(2)$
transformations:
\begin{equation}
{\bf x} \to {\bf x'} = {\bf S} {\bf x} {\bf S}^\dagger,
\end{equation}
where
\begin{equation}\label{fundamental}
{\bf S} = e^{{i\over2}\sigma_i ( \omega^i - i\nu^i)}.
\end{equation}
We have replaced the generators $N_i$ with Pauli matrices $\sigma_i$.
This is the usual realization of the action of the Lorentz group
in Minkowski space. One of the orbits of this action is the
upper sheet of the two-sheeted hyperboloid - the Euclidean $AdS_3$
space, where Lorentz group acts by isometries. The boundary of this
hyperboloid is the projective light cone, on which Lorentz group
acts by conformal transformations.

\end{document}